\documentclass[12pt,a4paper]{article}
    \pdfoutput=1
    \usepackage[format=hang]{caption}
    \usepackage{epsfig,amssymb,amsmath,graphicx,subcaption,verbatim,hyperref,ulem,bm,float,tensor,tikz}
    \usepackage{blkarray}
    \usepackage{xcolor}
    \usepackage{tikz}
    \usetikzlibrary{decorations.markings}
    \usepackage{amsthm}

    \definecolor{BrickRed}{RGB}{203, 65, 84}
    \definecolor{Violet}{HTML}{EE82EE}
    \definecolor{OliveGreen}{HTML}{556B2F}
    \definecolor{RoyalBlue}{RGB}{25,41,88}
    
    \hypersetup{colorlinks=true, linkcolor=BrickRed, citecolor=Violet, filecolor=OliveGreen, urlcolor=RoyalBlue, filebordercolor={.8 .8 1}, urlbordercolor={.8 .8 0}}

    \newcommand{\bZ}{\mathbb{Z}}
    
    \newcommand{\be}{\begin{equation}}
    \newcommand{\ee}{\end{equation}}

    \font\mybb=msbm10 at 11pt
    \def\bb#1{\hbox{\mybb#1}}
    \def\bZ {\bb{Z}}

    \def\ket#1{\left|#1\right>}

    \newcommand{\news}{\setcounter{equation}{0}}
    \def\bea{\begin{eqnarray}}
    \def\eea{\end{eqnarray}}
    \numberwithin{equation}{section}
    \makeatletter
    \renewcommand*\env@matrix[1][\arraystretch]{
      \edef\arraystretch{#1}
      \hskip -\arraycolsep
      \let\@ifnextchar\new@ifnextchar
      \array{*\c@MaxMatrixCols c}}
    \makeatother
    \usepackage{color}
    
\begin{document}
    
    \title{\vskip -45pt
    \vskip 20pt
    {\bf {\large Ishibashi States, Topological Orders with Boundaries and Topological Entanglement Entropy}}\\[20pt]}
    \author{ { Jiaqi Lou, Ce Shen,  Ling-Yan Hung}\\[20pt]
    $^1$State Key Laboratory of Surface Physics\\ 
    Fudan University,\\
    200433 Shanghai, China\\
    $^2$Collaborative Innovation Center of Advanced Microstructures,\\
    210093 Nanjing, China\\
    $^3$Department of Physics and Center for Field Theory and Particle Physics,\\ Fudan University\\
    200433 Shanghai, China\\
    $^4$Institute for Nanoelectronic devices and Quantum computing,\\ Fudan University,\\ 200433 Shanghai , China\\
     }
    
    \date{\today}
    \maketitle
    \vskip 10pt
    
    \begin{abstract}
    
    In this paper, we study gapped edges/interfaces in a 2+1 dimensional bosonic topological order and investigate how the topological entanglement entropy is sensitive to them.
We present a detailed analysis of the Ishibashi states describing these edges/interfaces making use of the physics of anyon condensation in the context of Abelian Chern-Simons theory, which is then generalized to more non-Abelian theories whose edge RCFTs are known. Then we apply these results to computing the entanglement entropy of different topological orders. We consider cases where the system resides on a cylinder with gapped boundaries and that the entanglement cut is parallel to the boundary. We also consider cases where the entanglement cut coincides with the interface on a cylinder. In either cases, we find that the topological entanglement entropy is determined by the anyon condensation pattern that characterizes the interface/boundary. We note that conditions are imposed on some {\it non-universal} parameters in the edge theory to ensure existence of the conformal interface,  analogous to requiring rational ratios of radii of compact bosons. 
       \end{abstract}
    
    \vfill
    
    
    \section{Introduction}\news 
    
    There are many aspects to understanding a topological phase, and consistent boundary conditions contain important information about the topological phase itself. For example, it is known that a topological bulk system and its gapped boundaries are related in a very special way -- namely that the category describing the bulk can be understood as the center of the category describing the boundary.\cite{kitaev_models_2012}
        Moreover, boundary conditions of topological phases are rich with structures, both physical and mathematical. In 2+1 dimensions, topological or gapped boundary conditions are connected to modular invariants in 2d CFTs, to conformal boundary conditions and Ishibashi states, to Lagrangian sub-algebra in a tensor category and also to the phenomenon of anyon condensation, to name a few prominent connections.  It is thus a curiosity whether these structures are connected to the non-trivial long-range entanglement structure of the system, which is also known to be an important characterizing feature of the topological phase.
        
        There has been a lot of works studying topological entanglement of topological phases, both in the context of their continuous field theoretic description and their discrete lattice description \cite{levin_detecting_2006,kitaev_topological_2006}. More recently, these studies have been generalized to cases where the system contains physical boundaries \cite{Wang:2018edf,hung_ground_2015,lu_theory_2012}.   These studies are carried out in the context of discrete lattice descriptions of various topological phases-- namely the Dijkgraaf-Witten lattice gauge theories. It is found for example that there are very interesting values of the topological entanglement that is sensitive to the boundary condition. In the current paper, we will re-derive these results using the alternative description of these phases via Chern-Simons theories. To be precise, we will consider the extended Hilbert space furnished by edge modes localized at the entanglement cut, which are described by an appropriate (rational) CFT following from the bulk-boundary correspondence.  The entanglement entropy presents itself as the ``left-right'' entanglement of an appropriate Ishibashi state describing the gluing of the edge modes that recovers the gapped bulk. 
      
      Our paper is organized as follows. It is divided into two parts, section 2 and 3. 
      
     We start with a discussion of these gapped boundaries and interfaces based on the gluing of edge modes in section 2 using the physics of anyon condensation. 
     After a brief review of important aspects of anyon condensation, we begin with analyzing the case of Abelian Chern-Simons theories.
     In particular, we work out in detail the Ishibashi states corresponding to gapped edges and interfaces.  The important new ingredients compared to previous works such as \cite{Sakai:2008tt, PandoZayas:2014wsa, Das:2015oha, Brehm:2015lja,brehm_entanglement_2015,Gutperle:2015kmw,fliss_interface_2017} are that we provide a careful treatment of the matching of anyon charges across the interface using tools in anyon condensation. The anyon condensation picture allows us to generalize the results to include cases with ground state degeneracies, and also to non-Abelian systems. 

       We thus explore non-Abelian cases where a description of their edges are known in terms of various RCFTs. 
       This provides an alternative perspective to the discussion of Ishibashi states at an interface in the CFT literature (see for example \cite{petkova_generalised_2001} and \cite{brehm_entanglement_2015} and references therein) based on chiral symmetry breaking. From the perspective of the 3d TQFT bulk, we find that both chiral symmetry breaking and enhancement at the interface are equally natural, and both scenarios can be captured based on the picture of anyon condensation.     

 These results are then applied to computing entanglement entropy in the second half of the paper in section 3.
 We confirm prior results obtained based on lattice gauge theories \cite{chen_entanglement_2018, Shi:2018krj,Shi:2018bfb}, and also obtain new results of topological entanglement across an interface depending on the anyon condensation pattern.   
Specifically,  the branching coefficients that determine the decomposition of anyons under anyon condensation is key to the computation of the entanglement entropy across a non-trivial interface. 
        
 These results should shed new light into the interplay of entanglement and boundaries.

    \section{Interfaces and anyon condensation} \label{sec:interfaces_anyoncond}
    In this paper, we would like to discuss how topological entanglement entropy is sensitive to gapped interfaces and boundaries in 2+1 dimensions.
    To do so, we need a systematic description of these interfaces/boundaries. Physically, these objects can be understood in terms of anyon condensation. 
    
    Consider phases A and B separated by an interface. The simplest possibility that characterizes this interface is that anyons in A can condense to give B. 
    But one can also imagine that A and B is separated by a third topological phase C. Suppose the interface between A and C is characterized by A condensing to C, and similarly the interface between B and C characterized by B condensing to C. Then we can subsequently squeeze C into thinner and thinner slab, and ultimately fuse the A$\vert$C and C$\vert$B interfaces to obtain a new interface. Properties of this novel A$\vert$B interface, as we are going to show, can be completely understood in terms of the A$\vert$C and C$\vert$B interface. 
    Then it becomes clear that the most generic interface $I_{C_1\cdots C_n}(A|B)$ between A and B can be understood by the fusion of an entire series of interfaces:  
    \be
    I_{C_1\cdots C_n}(A|B) \leftrightarrow \textrm{A}|\textrm{C}_1|\cdots \textrm{C}_n |\textrm{B}, 
    \ee
    with each interface characterized by a ``direct'' anyon condensation between the neighbours. In the following, we will collect some useful facts about anyon condensation.
    
    Anyon condensation is not unlike the Higgs mechanism. The heuristic rules of anyon condensation has been discussed at length in \cite{bais_broken_2002,bais_condensate-induced_2009,bais_theory_2009} and summarized in \cite{hung_generalized_2015}. 
    A precise and mathematical description of anyon condensation can be formulated in various ways, although these descriptions are essentially equivalent : in category theory where topological orders correspond to modular tensor categories, anyon condensation is described in terms of a commutative symmetric Frobenius algebra in the category. In the special case where the resulting phase B is trivial, the collection of condensed anyons in A forms a Lagrangian subalgebra. \cite{kapustin_topological_2011,davydov_witt_2010, Barkeshli:2013yta, barkeshli_classification_2013, Barkeshli:2013yta,
     Kong:2013aya, fuchs2013bicategories, levin_protected_2013}
    These modular tensor categories also supply the data that defines rational conformal field theories (RCFT). 
  A CFT is characterized by its chiral algebra $\mathcal{V}$, where all operators fall into its representations. In the case of RCFT, there are only a finite number of non-trivial representations, and each corresponds to a topological sector or simple object in the modular tensor category.  Anyon condensation then corresponds to extending the chiral algebra by otherwise non-trivial sectors corresponding to the condensed anyons. Specifically, it is known to correspond to the notion of ``conformal embedding'' \cite{bais_condensate-induced_2009}. 
    A ``conformal embedding'' also implies relationships between characters between CFT's. (See for example \cite{Fuchs:1992nq}.)
     Alternatively, topological interfaces can also be understood as boundary conditions in the CFT, or as Ishibashi states. The construction and properties of these Ishibashi states are indeed related to ``conformal embeddings'' described above.
    In the current paper, we will make use of the CFT description of interfaces heavily, which comes in particularly convenient in computing topological entanglement entropy via the Ishibashi states. 
    \subsection{A collection of useful facts} \label{useful_facts}
    A complete description of anyon condensation is beyond the scope of the current paper. Here, we would collect the bare minimal set of results that would be relevant for the computation of topological entanglement in later sections. 
    
    Let us call the parent phase A and the condensed phase B. Topological sectors in $A$ are denoted by $a,b,c\cdots$, and those of B are denoted $i,j,k$. 
    
    \begin{enumerate}
    \item Among these anyons in A, some of them would be able to move freely across the interface, while others stuck within A. The former are ``unconfined'' in the condensed phase B, while the latter ``confined"
    \item Among those anyons in A that are unconfined, they are mapped to anyons in B as they cross the interface. The map between them is describable by a matrix $b_{ai}$. 
    Each entry $b_{ai}$ is a non-negative integer, describing the ``multiplicity'', or ``different ways'' an anyon $a$ gets mapped to $i$.  The matrix $b$ is often called the branching matrix. \cite{francesco_conformal_1997,fuchs_topological_2007,hung_generalized_2015, lan_gapped_2015}
    
    \item In particular, all anyon $c$ satisfying $b_{c0}\neq 0$ are the ``condensed'' anyons that are mapped to the trivial sector of phase B. These anyons have bosonic self-statistics. In the case where all condensed anyons have quantum dimension 1, they are closed under fusion among themselves. When their quantum dimension is greater than 1, there must exist some fusion channel where two condensed anyons fuse to a third condensed anyon i.e. $\exists c_3$, such that $N_{c_1 c_2}^{c_3} \neq 0$, where $c_{i}$ are sectors that have condensed. 
    
    \item This $b$ matrix can also be understood as describing how $a$ as a representation gets decomposed into $i$. It reduces to group representation decomposition under restriction of a group $G$ to a sub-group $K$ when the condensation is framed within the context of breaking of a gauge theory with gauge group $G$.  Relation of these matrix elements with decomposition of representations can be found in \cite{bais_condensate-induced_2009,hung_ground_2015,hung_generalized_2015}.    
    
    \item The matrix $b_{ai}$ commutes with the $S$ matrix. i.e
    \be
    \sum_i b_{ai} S^B_{ij}  =  \sum_b S^A_{ab} b_{bj}
    \ee
    We note that there is a set of simple rules for practically obtaining these $b$ matrices given the topological data that defines the phase before condensation, and the set of condensed anyons \cite{bais_broken_2002,bais_condensate-induced_2009,bais_theory_2009}.  We note that the $b$ matrices are generically {\it not invertible}, which can be viewed as a consequence of conservation of quantum dimension as an anyon in the parent phase decomposes into anyons in the condensed phase. Quantum dimensions satisfy some important relations. In particular, there is a reduction in quantum dimension for phase $A$ to condense to phase $B$. i.e. $D_A > D_B$. Other relations can be found for example in \cite{hung_generalized_2015}. We will not review them here. 
    
    \item From the perspective of chiral algebra $\mathcal{V}_B$ of the CFT corresponding to the condensed phase B, it is an extension of the chiral algebra $\mathcal{V}_A$ of phase A by operators in the sectors $\mathcal{V}_A^a$ where $a$ are those satisfying $b_{a0} \neq 0$. This is relatively well established where the condensed anyons correspond to ``simple currents'', with unit quantum dimension and a finite order under fusion. ie. $a^n = 1$ for some finite integer $n$ \cite{Fuchs:1992nq}.     Another context in which examples are well understood are WZW models characterized by affine Lie algebra. For embeddings $h \subset g$, the condensed phase B has affine symmetry $g$ while the ``parent phase'' A  has affine symmetry $h$. Representations $i$ of $g$ are automatically representations of $h$ under restriction where the decomposition of representations again can be described by a matrix $b_{ab}$ \cite{Fuchs:1992nq}. 
    In these cases, the characters then satisfy the following relation:
    \be \label{characters_relation}
    \chi_b^B = \sum_a b_{a b} \chi_a^A. 
    \ee
    We note that in the case of simple current condensation, these $b_{ai}$ for fixed $i$ is non-vanishing over $a$ that falls into the fusion orbit with the condensate $J$. The value $b_{ai}$ is equal to   the number of stabilizers among the simple current condensate $J$ satisfying $J \otimes a = a$ \cite{Fuchs:1999zi, hung_generalized_2015, bais_condensate-induced_2009}. 
   One very important point is that {\it the central charges are unchanged before and after anyon condensation.}
    
    \item It is thus an important point to note that as $A$ condenses to $B$, the chiral symmetry algebra is extended, or enhanced, while the topological order appears to be "broken". This point has been noted already in \cite{bais_broken_2002,bais_condensate-induced_2009,bais_theory_2009}.  Note that since the $b$-matrices are generically not invertible, these relations between the characters always take the form of a character in the condensed phase (with enhanced chiral symmetry algebra) being equal to a sum over characters in the parent phase (with a smaller chiral symmetry algebra), i.e. (\ref{characters_relation}) is satisfied in generic anyon condensation. 
    
    \item On a manifold with interfaces, generically, there would be non-contractible cycle(s), and thus ground state degeneracy. A systematic way to construct ground state basis would be to construct anyon loops winding around non-contractible cycles. In the presence of interfaces connecting multiple phases, only anyons that are unconfined under the relevant anyon condensation describing the interface could pass through the interface. Each anyon line that crosses an interface with $a$ on one side and $i$ on the other of the interface satisfying $b_{ai} =1 $ would contribute to an independent state. For $b_{ai}>1$ one can construct $b_{ai}$ orthogonal states with the same choice of $a$ and $i$ that crosses the interface. (We will illustrate that with an example concerning the topological order $D(S_3)$ in the examples section below. ) 
    Our computation of the entanglement entropy would be based on these basis states. 
    
    \item An interface between phase A and B can always be folded to become a physical boundary of the product phase A$\otimes \overline{\textrm{B}}$. (The over-line refers to taking the time-reverse of B. ) Therefore, each interface can be related to a Lagrangian subalgebra in the category corresponding to the product phase. The simple objects, or anyons contained in this Lagrangian subalgebra are labeled by a pair $(a,i)$ where $b_{ai}\neq 0$. They are essentially ``condensed'' at the boundary. 
    
    \item The above discussion concerns interfaces characterized by A condensing to B. Now let us generalize to the case where the interface is characterized by both A and B separated by an intermediate phase C with infinitesimal width. More complicated interfaces can be analyzed entirely analogously. As noted, each interface separating phases related by directly condensing one of them is characterized by a simple $b$ matrix.
    An anyon $a$ from $A$ would match to anyon $c$ in $C$ in $b^{A|C}_{ac}$ number of ways, and that $c$ would match to an anyon $b$ in $B$ in $b^{C|B}_{cb}$ number of ways. 
    Therefore as we construct independent basis states based on anyon line passing through the interface, the pair $(a,b)$ would supply $N(a;b)$ states where
    \be \label{basis_construction_junction}
    N(a;b) = \sum_{c \in C} b^{A|C}_{ac}b^{C|B}_{cb}.
    \ee
    To specify the precise basis state, we can therefore attach a subscript to each matching anyon pair 
    \be
    (a,b) \to  |(a,b)\rangle_{\alpha_c}, 
    \ee
    where 
    $1 \leq \alpha_c \leq b^{A|C}_{ac}b^{C|B}_{cb} $ for non-vanishing 
    \be
    b^{A|C}_{ac}b^{C|B}_{cb} \equiv N(a;c;b).
    \ee 
    A natural basis state of this interface is constructed as follows. 
    
    \begin{eqnarray} 
    & |(a,b)\rangle_{\alpha_c} \,\qquad &\textrm{for}\,~ D_C> D_{A},D_{B}   \nonumber  \\
    &| c\rangle = \sum\limits_{\substack{a\in A,\, b\in B, \\ \alpha_c\le N(a;c;b)}} |(a,b)\rangle_{\alpha_c}, \, \qquad &\textrm{for}\,~ D_{A}, D_{B}> D_C . \label{natural_basisABC}    
    \end{eqnarray}

    Comparing with \cite{Fuchs:1999zi, petkova_generalised_2001,brehm_entanglement_2015} where the interface preserves only a sub-algebra of those of phases $A$ and $B$, it corresponds to the case in which $C$ as a topological order condenses to $A$ and $B$ separately. In this case, these basis states $|(a,b)\rangle_{\alpha_c}$ correspond to the intertwiner state in equation (2.5) in \cite{brehm_entanglement_2015}, and the multiplicity is given precisely by the $b$ matrix.  
    We do not however impose the ``modular invariance'' condition. This is a generic feature in computing entanglement entropy in Chern-Simons theory where we specialize to the ``anyon'' eigenstate. It is found that the Ishibashi states are the appropriate states which has the correct information about fluxes threading the edges.
    
    \item These series of condensation pattern would lead to hierarchies of identities between characters. For example, suppose A condenses to C and B also condenses to C. 
    Then we have
    \be
    \chi_c^C= \sum_a b^{A|C}_{ac} \chi^A_{a} = \sum_b b^{C|B}_{cb} \chi^B_{b}. 
    \ee


    \end{enumerate}

    In the following, we illustrate examples of interfaces in both Abelian and non-Abelian theories. Particularly, we focus on models related to the quantum doubles $D(G)$, where $G$ is some group $G$.
    These models have well known lattice descriptions whose boundaries have been studied in detail  (See for example \cite{kitaev_fault-tolerant_2003, bombin_family_2008,beigi_quantum_2011,cong_topological_2016,Bullivant:2017qrv,Hu:2017faw}), and they provide independent checks of the computations and proposal we will discuss in the paper. 
    
    \subsection{Examples in Abelian Chern Simons theories}
    In this section we will set the stage for computations in later sections. Particularly, we will describe a class of Abelian Chern-Simons theories that correspond to the $\mathbb{Z}_N$ quantum double. Some details of these models are reviewed in the appendix \ref{CSreview}. Particularly, the action is given by equations (\ref{CSaction}) and (\ref{kmatrixZN}). We will construct explicitly examples of their gapped boundaries and interfaces, by identifying the relevant condensed anyons, and constructing a corresponding set of Ishibashi states that describes the gapped boundary. A review and an explanation of the notations used in the main text of the paper is relegated to the appendix \ref{CSreview}.

    \subsubsection{Gapped boundaries in a $\mathbb{Z}_N$ theory} \label{bZN}

    Recall that a gapped boundary is characterized by anyon condensation that takes the topological order A to the trivial phase.  The set of condensed anyons would form a so called Lagrangian subalgebra. This has been discussed in general in \cite{Bais:2002ny,Bais:2002pb, bais_condensate-induced_2009, Bais:2008xf,kapustin_topological_2011,
  barkeshli_classification_2013, Barkeshli:2013yta,
     Kong:2013aya, fuchs2013bicategories, levin_protected_2013, wang_boundary_2015} and in the special case of Abelian CS theories, in \cite{kapustin_topological_2011,levin_protected_2013}.  In a $D(Z_N)$ quantum double, all such Lagrangian subalgebras are known. We can take the set $L$ of condensed anyons as 
    \be
    L = \{(P^{1}, P^{2})\},
    \ee
    where $(P^{1}, P^{2})$ is the pair of quantized quantum numbers (see equation (\ref{left_rightmodes}, \ref{left_right_momenta})) of the condensed sector.
    
    Among them there are two sets of gapped boundaries that are shared by all 
    $D(Z_N)$ theories and we will take them as examples for illustration purpose. These boundaries are called ``electric'' and ``magnetic'' boundaries respectively. Physically, the former correspond to the condensation of all electric charges and magnetic charges respectively i.e.
    \be \label{electricL}
    L_E = \{( N n + a, 0)\}, \qquad n \in \bZ, \qquad 0\leq a\leq N-1,
    \ee
    and similarly
    \be \label{magneticL}
    L_M = \{( 0, N m + b)\}, \qquad m \in \bZ, \qquad 0\leq b\leq N-1.
    \ee
    We note that these vectors that are collected into the condensed set $L$ are more selective than picking all charge vectors corresponding to the condensed topological sector. 
    In particular, they are ``self-null'' and ``mutually-null'' vectors satisfying 
    \be \label{null_condition}
    P_{l_i}^I K^{-1}_{IJ} P_{l_j}^J = 0, \qquad  \forall (P^1_{l_i}, P^2_{l_i}) \in L_{c},
    \ee
    where $L_c$ denotes a generic collection of vectors of condensates in a Lagrangian subalgebra.  This has been discussed at length in \cite{levin_protected_2013,wang_boundary_2015}, particularly how they are related to existence of corresponding relevant operators that could gap these edge modes. 
    
    Alternatively, one can think of these Lagrangian subalgebra as characterizing conformal boundary conditions \cite{cardy_boundary_1989,cardy_boundary_2004}. We note that the boundary theory has a set of  $U(1)$ global symmetries extended to a $U(1)$ Kac-Moody algebra.  The conserved currents are given by
    \be
    J^I_{x} = \frac{K^{IJ}}{2\pi} \partial_x \Phi_J. 
    \ee
    
 This implies that the zero mode of the current is given by
 \be \label{J0}
 J^I_{x,0} \equiv \int_0^l dx \, J^I_{x} = P^I.
 \ee   
    The Lagrangian subalgebra defines a boundary condition, or alternatively a boundary state $|\psi\rangle\rangle$ that preserves the following symmetries
    \begin{equation}\label{BC}
    P_{l_i}^I K^{-1}_{IJ} J^J_x |\psi\rangle\rangle=0.
    \end{equation}
    Using (\ref{J0}), this implies that the boundary condition is allowing the state to carry non-trivial expectation values of $P_{I_i}$ simultaneously if they are mutually null, as described in (\ref{null_condition}).  
    Indeed we only need a minimal set of vectors $(P^1_{l_i}, P^2_{l_i})$ that are linearly independent to generate the entire $L_c$.
    In the case of the electric boundary, we need only the null vector $(1,0)$. i.e.
    \be \label{celectric}
    K^{-1}_{12} J^2_x  |\psi\rangle\rangle_E \equiv \frac{1}{2\pi}\partial_x \Phi_1 |\psi\rangle\rangle_E =0.
    \ee
    Similarly a magnetic boundary would amount to taking the condensate vector  $(0,1)$, leading to 
    \be \label{cmagnetic}
    \frac{1}{2\pi}\partial_x \Phi_2 |\psi\rangle\rangle_M =0.
    \ee
    Now in terms of the right and left moving fields, the above conditions on the boundary state can be re-written as
    \be \label{Jbcstate}
    (J_L  \pm J_R)|\psi\rangle\rangle_{E/M}=0, \qquad  J_{L,R} = \frac{1}{2\pi} \partial_x \Phi_{L,R},
    \ee
    where $\Phi_{L,R}$ are related to $\Phi_{1,2}$ by (\ref{left_rightmodes}). We immediately note that the above equations implies that the states $|\psi\rangle\rangle_{E/M}$ 
    are indeed conformal boundary states satisfying the conformal boundary condition,
    \be
    (L_n - \bar L_{-n}) |\psi\rangle\rangle_{E/M} = 0,
    \ee
    where $L_n$ are the Virasoro generators of the left-moving modes and $\bar L_m$ the corresponding generators of the right-moving modes. 
    This follows from the fact that the stress tensor can be expressed as 
    \be
    T = \pi J_L J_L, \qquad \bar T =\pi J_R J_R
    \ee 
    by the Sugawara construction. We note that the Hamiltonian $H$ in (\ref{Hamiltonian}) is indeed given by
    \be
    H = L_0 + \bar L_0 - \frac{c}{12}, \qquad  \textrm{where} \,\, c=1.
    \ee
    In terms of the mode expansion, 
    \be
    (\alpha_{{L},n}\pm \alpha_{{R},-n})|\psi\rangle\rangle_{E/M}=0
    \ee
    The corresponding boundary Ishibashi state has the following form:
    \begin{equation}
    |\psi\rangle\rangle_{E/M}= \exp(- \frac{2\pi\epsilon}{l} H) \exp(\mp \sum^{\infty}_{n=1}\frac{{1}}{{n}}\alpha_{{L},-n}\alpha_{{R},-n}))|P_L,P_R\rangle\rangle_{E/M},
    \end{equation}
    where 
    \be
    P_L = \mp P_R,
    \ee
    for electric and magnetic boundaries respectively. 
    The boundary state is not normalizable, and so $\exp(- \frac{2\pi\epsilon}{l} H)$ serves as a regularization, with $\epsilon$ infinitesimal. The parameter $l$ is the length of the circle. 
    The norm of this state is then given by
    \be
    \langle \langle \psi | \psi \rangle \rangle = \frac{q^\frac{P_L^2}{2}}{\eta(q)}, \qquad q=e^{\frac{-8\pi\epsilon}{l}}
    \ee
    
    This has been discussed for example in \cite{fliss_interface_2017}, although we would like to make the connection to anyon condensation more transparent in the current discussion.


    \subsubsection{The $D(\bZ_M)- D(\bZ_N)$ interface -- chiral symmetry breaking and enhancement} \label{MNinterface}

    Consider a cylinder with an interface in the middle between $D(\bZ_M)$ occupying the top half of the cylinder, and $D(\bZ_N)$ the bottom half. 
    At the interface, there would be a set of upper edge (u) fields $\Phi^{(u)}_I$ contributed by $D(\bZ_M)$, and similarly a lower edge (l) fields $\Phi^{(l)}_I$ contributed by  $D(\bZ_N)$.
    It is convenient to fold across the interface, taking the time reversal of the upper edge. Doing that, the $K$ matrices are correspondingly given by
    \begin{equation}
    K^{(u)}=\begin{pmatrix}
     0&M \\ 
     M&0 
    \end{pmatrix}
    , \qquad 
    K^{(l)}=-\begin{pmatrix}
     0&N \\ 
     N&0 
    \end{pmatrix}
    \end{equation}
    The minus sign exchanges left and right modes in the folded theory. 
    Here for concreteness, we take $M=mC$, $N=nC$ for some common factor $C$ of $M$ and $N$. 
    For simplicity, we consider $n$ and $m$ to be relatively prime i.e. $(n,m)=1$, and that $m>n$. 
    As explained, every interface admits a description in terms of anyon condensation. In the current situation, we will consider one obvious class of interface between $D(\bZ_M)- D(\bZ_N)$. i.e. We can consider anyon condensation of both phases $D(\bZ_{M,N})$ simultaneously down to $D(\bZ_{C})$. This can be achieved by condensing $C$ units of electric charges in $D(\bZ_{M})$ and  $D(\bZ_{N})$ separately. i.e. The condensates correspond to a collection $C_{l,u}$ of sectors labeled by 
    \be C_{l} = \{(P^{(l)\,\,1} = N a + s C, P^{(l)\,\,2}= N b)\}, \qquad C_{u} =  \{(P^{(u)\,\,1} = M a + q C, P^{(u)\,\,2}= M b)\}
    \ee
    where $s,q$ are integers satisfying $0\leq s\leq n$ and $0\leq q \leq m$, and $a\in \bZ$. These $P^I$ are zero modes as defined in (\ref{left_rightmodes}). Note that these $C_{l}$ and $C_{u}$ individually do not form a Lagrangian subgroup, since the condensed phase is non-trivial. Now, sectors that will remain unconfined are given by
    \begin{align}
    &U_{l}=  \{(P^{(l)\,\,1}= N a_1+d_1 , P^{(l)\,\,2} = N a_2+n b_1)\} , \\
    &U_{u}= \{(P^{(u)\,\,1}= M a_3+d_2, P^{(u)\,\,2} = M a_4+mb_2)\}, 
     \end{align}
     for all $ 0\leq b_i \leq C-1$,   $ 0\leq d_1\leq n-1$, $0\leq d_2\leq m-1$ and all $a_i\in \bZ$. 
     
    \begin{figure}[t]
        \centering
    \includegraphics[width=0.4\textwidth]{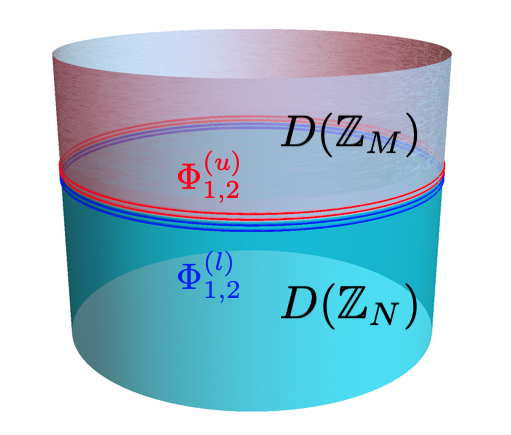}
    \label{fig:zmzn}
    \caption{The $D(\mathbb{Z}_N)-D(\mathbb{Z}_M)$ interface on a cylinder. The doubled red and blue lines at the interface denotes the pair of left and right moving modes in the upper and lower entanglement cut. }  
    \end{figure}

     In the condensed phase however, two sectors from the parent phases are identified if they are related by addition of vectors in $C_{l,u}$. Therefore, pairs of anyon lines from the lower and upper half of the cylinder can match up in the (infinitesimally thin) intermediate condensed phase if they are mapped to the same sector there. 
    This leads to the matching of quantum numbers given by
    \begin{align}
    &(P^{(l)\,\,1} = Na_1 + s C + x  \leftrightarrow P^{(u)\,\,1}= M a_3 + q C + x), \nonumber \\
    &\qquad\qquad\qquad\qquad \textrm{where} \,\,0\leq x\leq C-1,  \,\,  0\leq s \leq n-1,\,\, 0\leq q\leq m-1,  \label{matching1} \\
    &(P^{(l)\,\,2} = Na_2 + n b  \leftrightarrow P^{(u)\,\,2}= M a_4+  m b ), \,\, \textrm{where} \,\, 0\leq b\leq C-1 \label{matching2}
    \end{align} 
    One can see that $x,b$ are parametrizing the electric and magnetic sectors of $D(\bZ_C)$ respectively. 
    Since the intermediate $D(\bZ_C) $ phase is squeezed to infinitesimal size, this matching pattern must correspond to a matching condition of the fields that ultimately is identified with a conformal boundary condition appropriate for a gapped interface. One can check that
    \be \label{gluing_interface}
    n \partial_x \Phi^{(l)}_2 - m \partial_x \Phi^{(u)}_2 = 0, \qquad \partial_x \Phi^{(l)}_1 - \partial_x \Phi^{(u)}_1 =0
    \ee 
    recovers the matching of the sectors that is described in (\ref{matching1},\ref{matching2}).  Moreover, in terms of the overall $K$ matrix of the folded theory $\mathbb{K} = K^{(u)} \oplus K^{(l)}$, the above boundary conditions defines two linearly independent 4-component basis vectors $l_{1,2}$, 
    \be
    l^T_1 = (0,-m,0,n), \qquad l^T_2  = (-1,0,1,0)
    \ee 
    that satisfy the ``self-null'' and ``mutually-null'' conditions (\ref{null_condition}) and generate a Lagrangian subalgebra in the $\mathbb{K}$ theory. 
    We would like to determine if the matching condition (\ref{gluing_interface}) implies a set of conformal boundary conditions. One readily checks that imposing the above conditions do not ensure matching of $H^{(u)}$ and $H^{(l)}$ even for the zero mode contribution.  This can be attributed to the fact that taking the convenient choice for the velocity parameters so that $r=1$ in (\ref{left_rightmodes}), the matching condition (\ref{gluing_interface}) maps left-moving modes from one edge to both left and right moving modes on the other edge.  To recover conformal boundary conditions at the interface, we restore the parameters $r$ in  (\ref{left_rightmodes}), and demand that (\ref{gluing_interface}) relates left-moving modes only to right-moving modes. Denoting these parameters on the upper/lower edge by $r^{(u,l)}$, we have
    \be \label{ratioofr}
    \frac{r^{(u)}}{r^{(l)}} = \frac{m}{n}, 
    \ee
    so that (\ref{gluing_interface}) reduces to 
    \be
    \partial_x (\Phi^{(u)}_{L,R}- \Phi^{(l)}_{R,L})=0.
    \ee
    
    We note that (\ref{ratioofr}) is the analogue of the constraint on the radius of free bosons to obtain a topological interface in \cite{Bachas:2007td}. 
     The relation (\ref{ratioofr}) does not determine the velocity matrix uniquely. For convenience, we therefore set  $V_{11}^{(u)}=V_{11}^{(l)}=1$, $V_{22}^{(u)}=M^2$ and $V_{22}^{(l)}=N^2$. The constraints on $P^{(u,l)\,\,1,2}$ satisfying (\ref{gluing_interface}) can be summarized as follows:
     \be
    \frac{n}{N} P^{(l)\,\,1} \equiv (n a_1 + \frac{s C + x^1_l }{C} ) = \frac{m}{M} P^{(u) \,\,1} \equiv (m a_3 + \frac{q C+ x^1_u}{C}).
     \ee
    Similarly,
    \be
    \frac{P^{(l)\,\,2}}{N} \equiv (a_2 + \frac{n b_l}{N}) = \frac{P^{(u)\,\,2}}{M} \equiv (a_4+ \frac{m b_u}{M}) .
    \ee
    Finally, we thus have
    \begin{align}
    &x^1_l= x^1_u = x, \qquad n a_1 + s = m a_3 + q,  \label{electricinterface} \\
    &b_l = b_u \equiv b, \qquad a_2 = a_4 \equiv a.  \label{magneticinterface}
    \end{align}
    In (\ref{electricinterface}), each set of $(s,q,x,b)$ parametrizes a different charge sector in the upper and lower edge modes respectively.
    For fixed $(s,q,x,b)$, we can reparemtrize by
    \be
    n a_1 + s= m a_3 + q = m n \alpha + \beta,
    \ee
    where $0\leq \beta \leq m n - 1$ and for $\alpha \in \bZ$. For every $x$, there are $m*n$ distinct anyon pairs that can be joined at the interface. That is now completely parametrized by $\beta$.
    
    The Ishibashi state for a set of eigenvalues $s,q,x,b$ can now be written as
    \be \label{MN_ishibashi}
    |s,q,x,b \rangle\rangle =  e^{- \frac{2\pi\epsilon}{l} (H^{(u)}+ H^{(l)})}  \sum_{\alpha, a}\exp(\sum_\sigma \frac{1}{\sigma} \alpha^{(u)}_{L,{-\sigma}} \alpha^{(l)}_{R,{-\sigma}}) \, \exp(\sum_v \frac{1}{v} \alpha^{(u)}_{R,{-v}} \alpha^{(l)}_{L,{-v}}) | P^{(l)\,\,}_{L,R}= P^{(u)\,\,}_{R,L} \rangle, 
    \ee
    for $P^{(u,l)}_{L,R}$ related to the parameters defined in (\ref{electricinterface}, \ref{magneticinterface}) via (\ref{left_rightmodes}),
    and where $H^{(u,l)}$ are defined as in (\ref{Hamiltonian}). 
    
    The norm of this state is given by
    \be \label{define_rsxb}
    \langle\langle s,q,x,b| s,q,x,b\rangle\rangle \equiv \chi_{s,q,x,b}(\tau) = \sum_{\alpha,a}\frac{q^{\frac{(nmC\alpha+C\beta+x)^2+(a+\frac{n}{N}b)^2}{2}}}{\eta(q)^2},\qquad q\equiv e^{2\pi i \tau}=e^{-\frac{8\pi\epsilon}{l}}
    \ee
    We note that this function $\chi_{s,q,x,b}(\tau)$ is related to the corresponding characters of $D(\bZ_C)$ with electric and magnetic quantum numbers labeled by $(x,b)$
    \be \label{eq:character_ZC}
    \chi^{\bZ_C}_{x,b}(\tau)= \sum_{\substack{q \in \{0,\cdots m-1\} \\ s \in \{0,\cdots n-1\}}}\,\,\chi_{s,q,x,b}(\tau).
    \ee   
    
    One wonders what these characters $\chi_{s,q,x,b}$ are. 
    A moment's thought reveals that these are characters of sectors belonging to $D(\bZ_{Cnm})$ that remain unconfined as it condenses to $D(\bZ_{N})$ and $D(\bZ_{M})$. 
    In other words, when we insist upon preserving the resolution of individual anyons in A and B, our boundary condition (\ref{gluing_interface}) has invoked an intermediate topological phase $D(\bZ_{Cnm})$.  The Ishibashi state $|s,q,x,b\rangle\rangle$ describes chiral symmetry breaking of the $D(\bZ_{N})$ and $D(\bZ_{M})$ edge CFT's which has {\it fewer} sectors corresponding to the trivial sectors. As already described in the introduction, breaking of chiral symmetry is an enhancement of {\it topological symmetries} of the topological order i.e. There are more topological charges conserved individually. 
    
 What about the anyon condensation to $D(\bZ_C)$ that inspired the analysis of this section? 
 This is encoded in (\ref{eq:character_ZC}).  The intermediate phase $D(\bZ_C)$ has an enhanced chiral symmetry but diminished {\it topological symmetry}. All sectors labeled by different $s,q$ for fixed $x=b=0$ are all absorbed into its chiral symmetry algebra. Therefore the final interface described by an auxiliary $D(\bZ_C)$ with fixed $\bZ_C$ charges $x,b$ threading the interface is given by the Ishibashi state
 \be \label{eq:bZCishi}
 |x,b\rangle\rangle_{\bZ_C} = \sum_{s,q} |s,q,x,b\rangle\rangle.
 \ee
 
 To summarize, the interface $D(\bZ_N)-D(\bZ_M)$ described by the anyon condensation pattern of these phases to $D(\bZ_C)$ led to the Ishibashi state (\ref{eq:bZCishi}).
 In the process, we discovered another interface described by an intermediate phase $D(\bZ_{Cnm})$. It has an enhanced topological symmetry but diminished chiral symmetry, being related to $D(\bZ_{N,M})$ by condensing separately to each phase. This auxiliary phase $D(\bZ_{Cnm})$ is related to $D(\bZ_C)$ also by anyon condensation. 
 We note that  equation (\ref{eq:character_ZC}) is precisely the special case of (\ref{characters_relation}) describing the anyon condensation pattern that takes $D(\bZ_{Cnm}) \to D(\bZ_{C})$.
    
In other words, we have illustrated in explicit examples the extension and breaking of the chiral symmetry algebra by simple currents at an interface. A state where we sum over the sectors on either side of the interface corresponding to a single sector in $D(\bZ_{C})$ is a state that preserves a larger chiral symmetry algebra. This larger chiral symmetry algebra is characterized by the condensed phase  $D(\bZ_{C})$.
    On the other hand a boundary condition that preserves the individual identity of anyons in the two topological phases at the interface breaks the chiral symmetry algebra to a common subalgebra of the two phases, characterized by the topological order $D(\bZ_{Cnm})$. 
    
    In general, we can consider interfaces that preserve an even smaller chiral symmetry algebra, characterized by an even more complicated auxilliary topological order. It would then become possible that the same $(a,b)$ anyon junction at the interface has more than one independent way of joining, a scenario that we have briefly described near (\ref{basis_construction_junction}).
    
    This can be compared with the discussion  on CFT interfaces \cite{brehm_entanglement_2015}, which is characterized by chiral symmetry breaking. Here, from the perspective of topological phases, this has an immediate interpretation in terms of introducing an auxiliary topological phase that condenses in different ways to the phases A and B sandwiching the interface.
   In the discussion of topological orders however, it is equally natural to discuss chiral symmetry enhancement based on condensation of A and B itself. This thus complements the picture offered by the CFT analysis.
    
    %
    

    \subsection{Non-Abelian examples:  $D(S_3)$}
    
    In the following, we will present examples of gapped boundaries and interfaces involving non-Abelian theories.
    The example we will discuss in some detail involves the $D(S_3)$ model. It is basically an $S_3$ lattice gauge theory.
    The quantum double $D(S_3)$ of permutation group $S_3=\left< x,y|x^3=y^2=e,xyxy=e\right>$. $D(S_3)$ has 8 anyon types $\{A,B,C,D,E,F,G,H\} $. 
    It is known that $A$ is the trivial sector and $B$ the electric charge corresponding to the 1d representation of $S_3$. All the other anyons are non-abelian anyons with quantum dimension $>1$, and the total quantum dimension of the theory is $D=6$.
    
    A summary of the basic topological data of the model is relegated to the appendix \ref{S3model_review}.

    We would like to use the topological data to recover the corresponding edge states in the continuum limit. It is known that the set of topological data coincides with an orbifold CFT \cite{dijkgraaf_operator_1989}. Presumably the orbifold CFT recovers the correct edge CFT. In the case of Abelian theories $D(\bZ_N)$ where the edge CFT can be constructed explicitly via the Chern-Simons theory as reviewed at length in the appendix, we can see that the orbifold description is well founded.  
    
    The orbifold description is based on the determination of the chiral symmetry algebra of a chiral CFT. In \cite{dijkgraaf_operator_1989} it is assumed that group action is completely chiral, so that one can consider the holomorphic and anti-holomorphic algebra separately, and subsequently built up a diagonal modular invariant. As we have seen in the edge theory of the Chern-Simons description of the Abelian quantum doubles $D(\bZ_N)$, this is not the case in the edge theory. The left and right moving modes combine in particular ways to recover the electric and magnetic sectors. It is not the case that all the anyons of the quantum double gets completely encoded within the holomorphic sector alone.  The actual edge theory however, has to be a non-chiral theory where only a condensable set of anyons could generate modular invariants. 
    
    The analysis of \cite{dijkgraaf_operator_1989} can be done directly on the combined holomorphic and anti-holomorphic sector. i.e. Consider a chiral and anti-chiral symmetry algebra $\mathcal{C}= \mathcal{V} \otimes \bar{\mathcal{U}}$. In principle $\mathcal{U}$ may not be equal to $\mathcal{V}$. However, from the experience with explicit study of the edge modes of the $D(\bZ_N)$ model, it appears that indeed there is nothing to distinguish the left and the right, and $\mathcal{U} = \mathcal{V}$. The chiral and anti-chiral symmetry algebra must each contain a copy of the Virasoro algebra. As a non-chiral theory the holomorphic and anti-holomorphic central charges must be equal, and this is automatically satisfied.  
    
    Now for entirely the same reasons as described in \cite{dijkgraaf_operator_1989}, we expect that $\mathcal{C}$ enjoys an internal $S_3$ symmetry, and so it can be decomposed into 
    \be \label{chiral_symmetry}
    \mathcal{C}= \oplus_\alpha [[\phi_\alpha]] \otimes r_\alpha,
    \ee
    where $r_\alpha$ denotes representations of $S_3$.  The $[[\phi_\alpha]]$ labels a pair of left and right primary representations. Note that from experience with the Abelian case, it does not necessarily satisfy $h^\alpha_L = h^\alpha_R$, where $h^\alpha_{L (R)}$ are the (anti)-holomorphic  conformal dimensions. However, since the spin of a sector is given by $h_L-h_R$, we expect that the bosonic sectors, namely the pure electric (labelled by group representations) and pure magnetic (labelled by conjugacy classes) sectors in fact are left right symmetric. This expectation is borne out clearly in the Abelian case. In the orbifold CFT, only 
    \be
    \mathcal{C}_0 = [[\phi_e]] \otimes e
    \ee
    remains as the chiral symmetry.  The decomposition of $\mathcal{C}$ recovers the electric sectors. The magnetic and dyonic sectors come from the twisted sectors that generate non-trivial monodromies on the electric sectors. They are local only wrt $\mathcal{C}_0$. They can be classified in precisely the manner described in \cite{dijkgraaf_operator_1989}. 
    For completeness we enlist the twisted sector $\mathcal{T}$ here, which can be further decomposed into sub-sectors:
    \be
    \mathcal{T} = \oplus_{g, \alpha_g} [[\phi^g_{\alpha_g}]] \otimes r_{\alpha_g}.
    \ee
    Here, $g$ is a group element belonging to the group $G=S_3$ and $r_{\alpha_g}$ the representation of the centralizer group  $N_g$ of $g$ i.e. $[h,g]=1$ for $h\in N_g$.  While these twist fields have non-trivial monodromies with sectors in $\mathcal{C}$, these monodromies are only defined up to conjugation. Therefore $[[\phi^g_{\alpha_g}]] \otimes r_{\alpha_g}$ are entirely isomorphic for all $g \in A$ for some conjugacy class $A$. They are thus directly denoted as $[[\phi^A_{\alpha_{A}}]]\otimes r_{\alpha_A}$.   Characters for a sector is then defined as
    \be
    \chi^g_{\alpha_g}(q,\bar q) = \textrm{Tr}_{[[\phi^g_{\alpha_g}]]} [ q^{L_0- c/24} \bar{q}^{ \bar{L}_0- c/24}].
    \ee
    
    These distinct sectors are in one-to-one correspondence with the representations of $D(S_3)$ which are summarized in the appendix. 
    \subsubsection{Gapped boundaries of $D(S_3)$}
    In the following, we would like to understand the gapped boundaries of $D(S_3)$ from the perspective of the edge CFT.
    The set of gapped boundaries are again describable by anyon condensation. It is realized however that there is an efficient way of describing them in the lattice model. Namely, in the lattice realization as a $G=S_3$ lattice gauge theory, where the Hilbert space at each site is a $|G|= |S_3|= 6$ dimensional space labelled by the group elements of the group, the boundary sites can be restricted to span only a subgroup $K \subset G= S_3$ \cite{beigi_quantum_2011}.
     Each subgroup $K$ in fact corresponds to a particular set of anyon condensation at the boundary.  By computing overlaps of characters defined for the quantum double, these condensates, in fact the values of $b_{c0}$, the branching matrix introduced in the section \ref{useful_facts},  can be worked out efficiently \cite{beigi_quantum_2011}, giving an alternative route to the methods discussed in \cite{bais_broken_2002,bais_condensate-induced_2009,bais_theory_2009}. 
    
    We denote by $\mathcal{C}(K)$ the anyons that can condense at the boundary with boundary subgroup $K$. By calculating the inner product of characters one obtains the following table of condensed anyons associated with the respective boundary condition.

    \begin{table}[h]
        \centering
    \begin{tabular}{c|c}
        \hline\hline
        subgroup $K$ & $\mathcal{C}(K)= \oplus_{c} \, b_{c0} \,c$ \\
        \hline
        $\{e\} $ & $A\oplus B\oplus 2C$ \\
        $\{e,y\}$, $\{e,xy\}$, $\{e,x^2y\}$ & $A\oplus C\oplus D$ \\
        $\{e,x,x^2\}$ & $A\oplus B\oplus 2F$ \\
        $S_3$ & $A\oplus D\oplus F$ \\
        \hline\hline
    \end{tabular}
    \caption{Condensed anyons corresponding to boundary subgroup $K$}
    \label{tab:condense}
    \end{table}

    As we have mentioned in section \ref{useful_facts}, anyon condensation corresponds to extension of the chiral symmetry algebra. 
    For some simple cases, it appears fairly clear how the algebra extension should proceed. We will inspect some of them in detail in the following.

    {\bf \underline{1. $\mathcal{C}(K) = A\oplus B\oplus 2C$ }}
    
    In this case, it corresponds to taking the entirety of $\mathcal{C}$ defined in equation (\ref{chiral_symmetry}) to be the chiral symmetry algebra.
    As a result we can only keep sectors that have trivial monodromy with it, so that all the twisted sectors are discarded in the extended CFT. 
    The explicit way we are extending the chiral symmetry algebra also ensures that (\ref{characters_relation}) would hold. 
    Since $A,B,C$ all have trivial twists, they should satisfy $h_L= h_R$. The condensate should define conformal boundary states.
    We expect that a conformal boundary state $|B_e\rangle\rangle$ can be constructed by demanding
    \be \label{nonAbelianbcstate}
    L_n- \bar{L}_{-n}|B_e\rangle\rangle = 0, \qquad  (\mathcal{O}^L_{h^\alpha}(x) \mathcal{O}^R_{h^\alpha}(x))\vert_{t=0} |B_e\rangle\rangle =  |B_e\rangle\rangle 
    \ee
    where $O_{h^\alpha}^L(x)O_{h^\alpha}^R(x) \in [[\phi_{\alpha}]]$.  The second equation is the ``exponentiation'' of the condition (\ref{Jbcstate}) that appears in the 
    Abelian case, since $J^{L,R}\sim \partial_x\Phi_{L,R}$ and the exponentiation gives $\sim \exp( \Phi_{L}) \exp(\Phi_{R})$, which is the vertex operator corresponding to the condensed sector.  
    \\
    
    {\bf \underline{2. $\mathcal{C}(K) = A\oplus C \oplus D$ }}
    
    This is a more interesting case because $C$ is the 2d irrep  (electric charge) sector of $S_3$ and $D$ the twist (magnetic charge) sector corresponding to the conjugacy class of 2 cycles in $S_3$. This conjugacy class has 3 members (the three distinct 2 cycles), and therefore $D$ has quantum dimension equals 3. 
    However, according to the branching matrix and the identity (\ref{characters_relation}) expected to hold, it is suggesting that only part of the operator algebra corresponding to $C$ and $D$  participates in extending the chiral symmetry algebra.
    In these cases where the quantum dimension is integral and that they are in direct correspondence with representations and group elements of a discrete group, it is relatively straightforward to guess the precise form of the extension. 
    Namely, consider the sector $[[\phi_{\alpha=2}]] \otimes r_2$, where $r_2$ is a 2d vector space on which the 2d irrep acts. With this tensor product structure,
    we can easily decompose this space into 
    \be
    [[\phi_{2}]] \otimes r_2 = ([[\phi_2]] \otimes v_1 )\oplus ( [[\phi_2]] \otimes v_1), \qquad r_2 = v_1 \oplus v_2.
    \ee
    How should this basis $v_i$ be chosen? Looking ahead that all operators in the chiral symmetry algebra are mutually local, we look into one of the three 2-cycles of $S_3$ -- say (12). In the 2d rep, there exists a basis in which its representation matrix is diagonalized, with eigenvalues $\pm1$ with corresponding eigenvectors $v^{(12)}_{\pm}$. We will take $v_{1,2} = v^{(12)}_{\pm}$ respectively. Altogether, we come up with the following extension of the chiral symmetry algebra
    \be
    \mathcal{C}^{\textrm{ext}} = ([[\phi_e]]\otimes e )\oplus( [[\phi_2]]\otimes v^{(12)}_+ )\oplus ([[\phi^{g=(12)}_{e}]]\otimes e).
    \ee 
    This ensures that the operators in $\mathcal{C}^{\textrm{ext}}$ are all mutually local. Again we note that the extension is consistent with the identity (\ref{characters_relation}).
    Corresponding boundary states should be constructed in an analogous manner as in (\ref{nonAbelianbcstate}). 
    
    The other cases can be analyzed in an analogous manner as in cases 1,2 above.

    \subsubsection{The $ D(\bZ_4)-D(S_3)$ interface} \label{43_interface}
    
    We would like to explore a $D(\bZ_4)-D(S_3)$ interface. It is known that both $D(\bZ_4)$ and $D(S_3)$ can condense into a $D(\bZ_2)$ phase. 
    
    In $D(\bZ_4)$, it is two units of the electric charge that condenses. The unit charge $e$ and $3e$ are both identified with the $\bZ_2$ unit charge, and two units of the magnetic charge identified with the unit magnetic charge in $\bZ_2$. It is thus characterized by the following non-vanishing $b$ matrix elements.
    \begin{align}
    & b^{4|2}_{(0e,0m),(0e,0m)} = b^{4|2}_{(2e,0m),(0e,0m)}= 1; \\
    &b^{4|2}_{(1e, 2a m),(1e, a m)} = b^{4|2}_{(3e, 2a m),(1e, a m)} = 1, \qquad a = \{0,1\}. 
    \end{align}
    
    On the $D(S_3)$ side, the condensation is slightly more complicated. This is discussed to some detail in \cite{bais_condensate-induced_2009}. 
    We can summarize the condensation pattern by the following non-vanishing $b$ matrix elements
    \begin{align}
    &b^{S_3|2}_{A, (0e,0m)} = b^{S_3|2}_{C, (0e,0m)} = 1; \\
    & b^{S_3|2}_{B, (1e,0m)}= b^{S_3|2}_{C, (1e,0m)} = 1;\\
    &  b^{S_3|2}_{D, (0e,1m)}= 1;\\
    &  b^{S_3|2}_{E, (1e,1m)} =1.
    \end{align}
    Each anyon in  $D(S_3)$ that remains unconfined in $D(\bZ_2)$ can now be connected to two different $D(\bZ_4)$ anyons. e.g. $A \to 0, 2e$. 
    
    A conformal boundary state that respects the chiral symmetry algebra of $D(\bZ_2)$ is obtained by a sum over basis states
    \be
    |c\in D(\bZ_2)\rangle = \sum_{\substack{a\in D(\bZ_4), b \in D(S_3),\\  \alpha_c \leq b^{4|2}_{a c} b^{2| S_3}_{cb} \neq 0}} |(a,b)\rangle_{c, \alpha_c}.
    \ee

    Alternatively, we can construct boundary conditions that breaks chiral symmetry algebra by embedding $D(\bZ_4)$ and $D(S_3)$ into a common larger topological phase.
    Simple possibilities include $D(S_3\times \bZ_4)$ or more interestingly, $D(S_4)$. 
     
     \begin{figure}[!b]
    	\centering
    	\begin{subfigure}{0.35\textwidth}
    		\includegraphics[width=\textwidth]{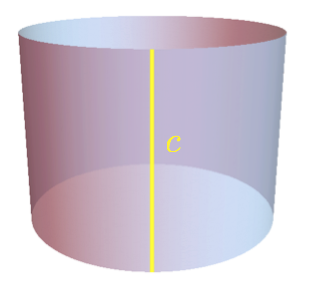}
    		\caption{anyon line basis}
    		\label{fig:wline}
    	\end{subfigure}
    	~ 
    	\begin{subfigure}{0.35\textwidth}
    		\includegraphics[width=\textwidth]{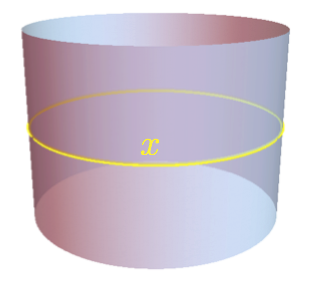}
    		\caption{anyon loop basis}
    		\label{fig:wloop}
    	\end{subfigure}
    	\label{fig:gs}
    	\caption{For a cylinder topology with non-trivial GSD, the ground states can be specified either by anyon line connecting the upper and lower physical boundary, or by anyon loop winding around non-contractible cycle. }
    \end{figure}

    \section{Entanglement of cylindrical regions in a cylinder} \label{EEcylinder}
    
    \subsection{A cylinder without interfaces}
    In this section, we will present examples computing entanglement of a cylindrical region $R$ embedded in a cylinder.

    A general solution is obtained in three steps. 
    \begin{figure}[!t]
        \centering
    \includegraphics[width=0.4\textwidth]{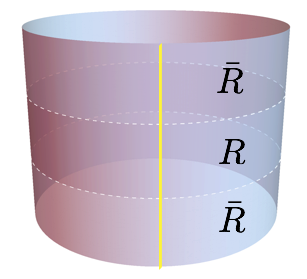}
    \label{fig:noface}
    \caption{A cylindrical region $R$ is embedded in a cylinder. The entanglement cut (white dashed line) separates region $R$ and $\bar{R}$. An anyon line (yellow) connects the upper and lower boundary.}  
    \end{figure}
    First, we construct ground state basis for given gapped boundaries of the cylinder. 
    Then for each of the orthogonal basis state we construct the appropriate Ishibashi state describing the entanglement cut.
    Finally, one can compute the entanglement across the entanglement cut. The strategy is basically identical to \cite{fliss_interface_2017} with the extra burden of determining the degenerate ground state. 
    As already discussed at length in the previous sections, a physical gapped boundary is characterized by anyon condensation such that the condensed phase becomes trivial with all anyons confined. The construction of ground state basis on a cylinder has been discussed at length in \cite{hung_ground_2015}. A convenient set of basis is obtained by attaching an anyon line that connects the upper and lower boundary.(Fig \ref{fig:wline})  Each such anyon must be a shared member of the condensates of the top and bottom boundaries. The number of independent states for 1 given anyon line can be greater than 1. In general {the number of independent basis state for a given anyon line $a$ is $n_a$,}\footnote{We have implicitly assumed that time reversal is an automorphism of the topological order and that the dual of an anyon $a$ is its time-reversed $\bar{a}$. They fuse to the trivial sector with exactly one unique fusion channel. i.e. $N_{a\bar{a}}^1 = 1$}
    \be
     n_a =  b^{\textrm{bottom}}_{a0} b^{\textrm{top}}_{a0}.
    \ee
    The total number of independent degenerate ground states on a cylinder is then 
    \be \label{cylinder_GSD}
    \textrm{GSD}_{\textrm{cylinder}} = \sum_a b^{\textrm{bottom}}_{a0} b^{\textrm{top}}_{a0}.
    \ee
 We have implicitly taken $N^{a}_{a0} =1$   For a given basis state constructed above, we consider the CFT state that describes the entanglement cut. In \cite{wen_edge_2016}, chiral topological orders are considered. As such each entanglement cut is described by an Ishibashi state that satisfies
    \be 
    (L_n - \bar{L}_{-n}) |h_a\rangle \rangle = 0, \qquad |h_a \rangle \rangle = \sum_{k} \exp(- \frac{2\pi}{l} \epsilon (L_0+ \bar{L}_0 - \frac{c}{12}) )|h_a, k\rangle_L \otimes |h_a,k\rangle_R 
    \ee
    where $h_a$ labels the primary sector with conformal dimension $h_a$ corresponding to the anyon line piercing through the entanglement cut. 
    We note that each cut creates an upper and lower edge, related to each other by time-reversal under a fold. For a chiral topological order, this gives rise to a pair of left and right moving modes, explaining the $L,R$ labels above. The conformal boundary condition is therefore a matching condition across the entanglement cut which is a trivial (non-existent) defect. 
    
    In the situations considered in the current paper, the topological orders are non-chiral. This is a necessary condition to ensure that the physical boundaries can be gapped at all. Therefore, the edge modes are already non-chiral. The upper and lower half edges at the entanglement cut thus make up two sets of $\{L^\sigma_n, \bar{L}^\sigma_n\}$, where we include a super-script $\sigma = \{l,u\}$ to denote the pair of Virasoro operators for the lower/upper edge at a cut. The matching condition satisfied by the generalized Ishibashi state should be enhanced to
    \be \label{eq:matchL}
    (L^l_n - \bar{L}^u_{-n})| h_a\rangle \rangle = (L^u_n - \bar{L}^l_{-n})| h_a\rangle \rangle = 0.
    \ee
    Similarly, for any other conserved currents we also expect them to satisfy a matching condition.
    \be \label{matchJ}
    (J^l_n - \bar{J}^u_{-n}) | h_a\rangle \rangle= (J^u_n - \bar{J}^l_{-n})| h_a\rangle \rangle = 0.
    \ee
    The generalized Ishibashi states satisfying the above condition is then given by
    \be \label{eq:cutishibashi}
    | a\rangle \rangle = \sum_{k,\bar{k}} \exp(-\frac{2\pi\epsilon}{l} H)| (h^L_a, k), (h^R_a, \bar{k} ) \rangle_{l} \otimes | (h^L_a, \bar{k}), (h^R_a, {k}) \rangle_{u}  
    \ee
    where the pair $(h^L_a, h^R_a)$ corresponds to the conformal dimension of the primary representation that is associated to the anyon, and $k,\bar{k}$ are the level number of the left and right moving modes on the same side of the edge at the entanglement cut. They are summed independently. We note also that $h^L_a$ is generically not equal to $h^R_a$. In the case of the Abelian Chern-Simons description of the $D(\bZ_N)$ theories, bound states of electric and magnetic charges correspond to $P_L \neq P_R$ in (\ref{left_right_momenta}). In those cases therefore $h^L_a \neq h^R_a$. Also the Hamiltonian $H$ here is defined as
    \be
    H = L_0^l + L_0^u + \bar{L}_0^l + \bar{L}_0^u - 2\frac{c+ \bar{c}}{24}, \qquad c= \bar c \,\, \textrm{for a non-chiral theory.}
    \ee
    The inner product is given by
    \be
    \langle\langle a | a \rangle\rangle =\chi_a(\tau),  \qquad  q\equiv \exp(2\pi i\tau) =e^{-\frac{8\pi\epsilon}{l}}
    \ee
    This factor $\chi_a$ is the character characterizing the anyon $a$. Note that in our examples where the topological order is non-chiral, these characters involve contributions from both the left and right moving sectors. 
    
    Putting together the two entanglement cuts, the state describing them is given by
    \be
    |\psi\rangle = | a\rangle \rangle_{\textrm{top cut}} \otimes |a\rangle \rangle_{\textrm{bottom cut}}.
    \ee
    The anyon characterizing the upper and the lower cut should be the same, given that the same anyon line cuts through both entanglement cuts. 
    
    The reduced density matrix is obtained by tracing out the upper half edge in the top entanglement cut, and the lower half edge in the bottom cut. The Renyi entropy is then given by \cite{wen_edge_2016,PandoZayas:2014wsa, fliss_interface_2017,brehm_entanglement_2015, Gutperle:2015kmw} (and also references therein)
    \be
    S^{(n)}=\frac{1+n}{n}\frac{\pi l (c+ \bar c)}{48\epsilon}-2 \text{ln}D+\frac{1}{1-n}\text{ln}\sum_a|\psi_a|^{2n}d_a^{2-2n}.
    \ee
    Finally, the entanglement entropy is obtained by taking the limit $n\to 1$, giving,
    \be \label{SEE_general}
    S_{EE}=\frac{\pi l  (c+ \bar c)}{24\epsilon}-2\ln D+2\sum_a|\psi_a|^2\text{ln}d_a-\sum_a|\psi_a|^2\text{ln}|\psi_a|^2.
    \ee
    
    We note that the current perspective based on anyon flux piercing the entanglement cut is closely related to the fusion-basis viewpoint described in \cite{Yidunyuting}.
    
    \subsection{Examples in Abelian Chern-Simons theory}
    
    In this section, we will illustrate the discussion above with explicit examples in a class of Abelian Chern-Simons theory, namely the field theoretic equivalence of the $D(Z_N)$ models reviewed in some detail in section \ref{CSreview}.  
    For later convenience, we label the top and bottom entanglement boundary by $b_1$ and $b_2$ respectively, while the top and bottom physical boundaries we denote by $B_1$ and $B_2$ respectively.\par
    \par
    Let us first construct the Ishibashi state at the entanglement cut $b_1$. The $K$ matrix that describes both the upper $(u)$ and lower $(l)$ edges at the cut are both corresponding to the $D(Z_N)$ theory. It is convenient to fold across the cut and take the time-reversal of the upper edge, so that
    we have:
    \begin{equation}
    K^{(l)}= - K^{(u)}=\begin{pmatrix}
     0&N \\ 
     N&0 
    \end{pmatrix}
    \end{equation}

    
    Here, the matching condition corresponding to (\ref{matchJ}) satisfied by the (generalized) Ishibashi state $|h_a\rangle\rangle$ reads 
    \begin{equation} \label{matchingsimple}
    \begin{cases}\partial_x(\Phi^{(l)}_L-\Phi^{(u)}_R )|h_a\rangle\rangle=0\\\partial_x(\Phi^{(l)}_R-\Phi^{(u)}_L )|h_a\rangle\rangle=0\end{cases}.
    \end{equation}
    This implies
    \begin{equation}
    \begin{cases}(\alpha^{(u)}_{L,n}-\alpha^{(l)}_{R,-n})|h_a\rangle\rangle=0\\(\alpha^{(u)}_{R,-n}-\alpha^{(l)}_{L,n})|h_a\rangle\rangle=0\end{cases}
    \end{equation}
    This gives
    
    \begin{equation}
    |h_a\rangle\rangle=\exp(\sum^{\infty}_{n=1}\frac{{1}}{{n}}\alpha^{(l)}_{{L},-n}\alpha^{(u)}_{{R},-n})\exp(\sum^{\infty}_{s=1}\frac{{1}}{{s}}\alpha^{(l)}_{{R},-s}\alpha^{(u)}_{{L},-s})|P^{(l)}_L,P^{(u)}_R\rangle\otimes|P^{(l)}_R,P^{(u)}_L\rangle
    \end{equation}
    
    We still need to determine the value of zero modes. Using (\ref{left_right_momenta}) and the matching conditions above, we have
    
    \begin{equation}
    \begin{aligned}[h]
    &P^{(u)}_{L,R}=P^{(l)}_{R,L} = \frac{1}{\sqrt{2N}} \bigg( \pm(N C_1+ c_1 )+ (N C_2+ c_2) \bigg)\\
    &C_1,C_2\in\mathbb{Z}\\
    &c_1, c_2\in[0,N-1]
    \end{aligned}
    \end{equation}
    
    The boundary conditions on $B_1$ and $B_2$ impose restrictions on the mod $N$ quantum numbers $c_{1,2}$ which determine the anyon line pierces through the edge .
    Consider the case where $B_1$ is characterized by the electric condensate, it implies $c_2=0$. If in addition we have $B_2$ characterized by the magnetic condensate, then  $c_1=0$. 
    We are then left with
    \begin{equation} \label{EM_gscharge}
    \begin{aligned}[h]
    &c_1=0, \qquad c_2=0\\
    &P^{(u)}_{L,R}=P^{(l)}_{R,L}=\sqrt{\frac{N}{2}} \bigg( \pm C_1 +  C_2 \bigg)
    \end{aligned}
    \end{equation}
    The vanishing of the $c_i$ quantum numbers means that there is only one unique anyon line that can pierce both the bottom and top physical boundaries, and thus the ground state of the CS theory on the cylinder with such boundary conditions is unique. All of these $C_i$ quantum numbers correspond to the same topological sector, and they would eventually be summed over in the Ishibashi state. 
    The state at $b_2$ can be carried out similarly and the full specification of the state at the entanglement cut is given by
    \begin{equation}
    \begin{aligned}[h]
    &|B\rangle\rangle={e^{-\epsilon (H_{b_1}+H_{b_2})}}|c_1=c_2=0\rangle\rangle_{b_1}\otimes|c_1=c_2=0\rangle\rangle_{b_2}\\
    &|c_1=c_2=0\rangle\rangle_{b_1}=\sum_{C_1,C_2}\exp(\sum^{\infty}_{n=1}\frac{{1}}{{n}}\alpha^{(u_1)}_{{L},-n}\alpha^{(l_1)}_{{R},-n})\exp(\sum^{\infty}_{s=1}\frac{{1}}{{s}}\alpha^{(u_1)}_{{R},-s}\alpha^{(l_1)}_{{L},-s})|P^{(u_1)}_L,\, P^{(l_1)}_R, \, P^{(u_1)}_R,\, P^{(l_1)}_L\rangle_{b_1}\\
    &|c_1=c_2=0\rangle\rangle_{b_2}=\sum_{C_3,C_4}\exp(\sum^{\infty}_{n=1}\frac{{1}}{{n}}\alpha^{(u_2)}_{{L},-n}\alpha^{(l_2)}_{{R},-n})\exp(\sum^{\infty}_{s=1}\frac{{1}}{{s}}\alpha^{(u_2)}_{{R},-s}\alpha^{(l_2)}_{{L},-s})|P^{(u_2)}_L,\,P^{(l_2)}_R, \,P^{(u_2)}_R,\,P^{(l_2)}_L\rangle_{b_2}
    \end{aligned}
    \end{equation}
    with zero modes
    \begin{equation}
       \label{eq:zero_modes}
    \begin{cases}
    P^{(u_{1})}_{L,R,b_1}=P^{(l_1)}_{R,L,b_1}=\sqrt{\frac{N}{2} }(\pm C_1+C_2)\\
    P^{(l_2)}_{L,R,b_2}=P^{(u_2)}_{R,L,b_2}=\sqrt{\frac{N}{2}}  (\pm C_3+ C_4)
    \end{cases}
    \end{equation}
    Clearly this is an unique ground state that consist of only one topological sector, which is consistent with the GSD of $Z_N$ on a cylinder with such kind of anyon condensation.
    
    \begin{equation}
    \begin{aligned}[h]
    N_B \equiv \langle\langle B| B\rangle\rangle &=\sum_{C_1,C_2,C_3,C_4}\frac{{e^{-\frac{{4\pi \epsilon}}{{l}}N((C_1)^2+(C_2)^2+(C_3)^2+(C_4)^2)}}}{{\eta(e^{-\frac{{8\pi \epsilon}}{{l}}})^4}}\\
    &=\frac{{1}}{{N^2\eta(e^{-\frac{{\pi l}}{{2\epsilon}}})^4}}\sum_{k_1,k_2,k_3,k_4}e^{-\frac{{\pi l (k_1^2+k_2^2+k_3^2+k_4^2)}}{{4N\epsilon}}}\\
    &\overset{l/\epsilon\rightarrow\infty}{\rightarrow}\frac{{q^{-1/6}}}{{N^2}}.
    \end{aligned}
    \end{equation}
    
    
    
    
    
    
    
    
    
    The full density matrix $\rho=N_B^{-1} |B\rangle\rangle\langle\langle B|$. After tracing out the the $u_1, l_2$ modes, we obtain the reduced density matrix $\rho_A$. Finally we have,
    \begin{equation}
      \label{eq:ee}
    \begin{aligned}[h]
    &\textrm{Tr}_L(\rho_A^n)\overset{l/\epsilon\rightarrow\infty}{\rightarrow}q^{1/6(n-\frac{1}{n})}N^{2(n-1)}\\
    &S_{EE}=-\underset{n\rightarrow1}{\lim}\partial_n \textrm{Tr}(\rho_L^n)=\frac{{\pi l}}{{6\epsilon}}-2\ln N
    \end{aligned}
    \end{equation}
    \par
    Other cases with different boundaries can be carried out with the same procedure. Supposed $B_1,B_2$ are both characterized by the electric condensates, then $c_2=0$ while $c_1$ can take any value. In which case, a generic state would look like 
    \be
    |\psi\rangle = \sum_{c_1} \psi_{c_1} |c_1\rangle\rangle_{b_1} \otimes |c_1\rangle\rangle_{b_2}.
    \ee
    The resulting entanglement entropy is given by
    \begin{equation}
    S_{EE}=\frac{{\pi l}}{{12\epsilon}}-2\ln N-\sum_a|\psi_{c_1}|^2\ln |\psi_{c_1}|^2.
    \end{equation}
    
    We note that the quantum dimension $D$ of the $\bZ_N$ models is precisely $D=N$.
    Therefore the above is recovering the classic result of topological entanglement entropy \cite{kitaev_topological_2006,levin_detecting_2006,dong_topological_2008,fliss_interface_2017}, in which each disconnected component of the entanglement cut contributes to $-\ln D$ and that the anyonic lines only contribute via  the  $|\psi_{c_1}|$ since $\ln d_c=0$ in Abelian theories.

    \subsection{The $D(S_3)$ model}
    
    Here we would like to illustrate with one explicit example of a non-Abelian theory, namely, the $D(S_3)$ quantum double. The physical gapped boundaries are again characterized by anyon condensation. The topological data of this model and the allowed gapped boundary conditions are reviewed in the appendix. 
    The construction of ground state basis states on a cylinder with gapped boundary conditions has been discussed at the beginning of this section. 
    
    For concreteness, we consider the top boundary to be characterized by the condensate $A\oplus B\oplus 2 C$ and the bottom boundary by $A\oplus C \oplus D$.
    The number of degenerate ground state is equal to 3, using (\ref{cylinder_GSD}). 
    The three basis states are respectively given by an $A$ line (denoted $|A\rangle$) or two orthogonal configurations of $C$ lines ($|C_{1,2}\rangle$) attached at the two ends of the cylinder.
    For a generic state, we have
    \be
    |\psi\rangle = \psi_A |A\rangle + \psi_{C_1} |C_1\rangle + \psi_{C_2}|C_2\rangle, \qquad |\psi_A|^2 +  |\psi_{C_1}|^2 + |\psi_{C_2}|^2  =1
    \ee
    
    We can use equation (\ref{SEE_general}) to compute the entanglement entropy of the cylindrical region. We note that both $|C_{1,2}\rangle$ would lead to the same characters when we compute $\textrm{Tr}\rho_{R}^n$. As a result in practice we can rewrite  
    \be
     \psi_{C_1} |C_1\rangle + \psi_{C_2}|C_2\rangle = \sqrt{|\psi_{C_1}|^2 + |\psi_{C_2}|^2} | \tilde C\rangle, \qquad  | \tilde C\rangle = \frac{1}{\sqrt{|\psi_{C_1}|^2 + |\psi_{C_2}|^2}} \left(\psi_{C_1} |C_1\rangle + \psi_{C_2}|C_2\rangle\right).
    \ee
    
    Finally, using topological data supplied in appendix \ref{S3model_review} and substituting into (\ref{SEE_general}), the topological entanglement entropy of a cylindrical region (subtracting the area term) not touching the boundaries is given by
    \begin{align} 
    &S_{EE} - \frac{\pi (c+\bar c) l}{12 \epsilon}  \\
    & = -2\ln 6 +2(|\psi_{C_1}|^2 + |\psi_{C_2}|^2)\ln 2 - \left(|\psi_A|^2|\ln |\psi_A|^2 + (|\psi_{C_1}|^2 + |\psi_{C_2}|^2 )\ln (|\psi_{C_1}|^2 + |\psi_{C_2}|^2)\right). \label{S3cylinderEE}
    \end{align}
    This result can be compared with \cite{chen_entanglement_2018}. It requires some work to decipher the basis in the lattice theory in terms of the anyon basis described above. The technical details are relegated to the appendix. We find that (\ref{S3cylinderEE}) is in complete agreement with the result in \cite{chen_entanglement_2018}.

    {\subsection{Comments on cylindrical regions $R$ containing an interface}
    
    Consider a cylinder consisting of two phases $X$ and $Y$, respectively occupying the top half and the bottom half of the cylinder connected by an interface. 
    The ground states are again characterized by anyon lines connecting the bottom physical boundary to the top half of the physical boundary, except that the topological sector of the anyon must change at the interface. The allowed matching anyon sectors at the interface is controlled by the $b$ matrix characterizing the boundary, as explained already in section \ref{useful_facts}.
    
    This, combining with a classification of gapped boundaries of each phase, gives a complete basis of degenerate ground state on the cylinder. i.e. the anyon pair that matches at the interface must also be ones that are allowed to end at the respective physical boundaries at the end of the cylinder. 
    The ground state degeneracy is given by
    \be
    \label{eq:gsd}
    \textrm{GSD}_{X|Y} = \sum_{a\in X,\mu\in Y,i \in C} b^{X|I}_{a 0} b^{X|C}_{ai} b^{C|Y}_{i\mu} b^{Y|I}_{\mu 0},
    \ee
    where $b^{A|B}$ denotes the $b$ matrix characterizing anyon-condensation between phases $A$ and $B$. We denote the trivial phase by $I$, and $C$ an auxilliary phase that characterizes the interface between $A,B$. 
    Suppose the anyon pair matched at the interface is labeled by $(a,\mu)$, then the entanglement entropy of the cylindrical region $R$ containing the interface -- i.e. where the interface does not touch the entanglement cut -- can be computed by an analogous formula as in (\ref{SEE_general}). This gives
    \be
    S_{EE}(a,\mu)_{X|Y} =  \frac{\pi l(c+\bar c)}{24\epsilon}-\ln D_X - \ln D_Y +(\ln d_a + \ln d_{\mu}),
    \ee
    where $D_{X,Y}$ denotes the quantum dimensions of the respective phases, and $d_{a,\mu}$ the quantum dimension of the respective anyons.

    \section{Entanglement across interfaces}
    
    In this section, we would like to consider entanglement entropy across an interface that is pierced by an anyon line. We will focus particularly on the entanglement that is contributed at the interface. 
    The interface can be described by a hybrid Ishibashi state that is described in detail in section \ref{sec:interfaces_anyoncond}. i.e. One introduces intermediate phase(s) $C_{(n)}$.
    For concreteness, we first consider the case where there is one intermediate phase $C$. 
    \begin{figure}[h]
        \centering
    \includegraphics[width=0.4\textwidth]{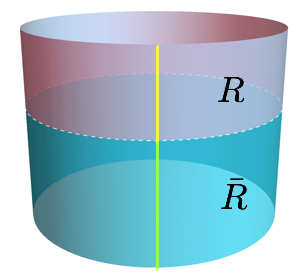}
    \label{fig:interface}
    \caption{There is an interface between region $R$ and $\bar{R}$, which coincides with the entanglement cut (white dashed line). Different anyons are matched across the interface.}   
    \end{figure}

    Taking the basis state described in (\ref{natural_basisABC}), and using the relation of the characters (\ref{characters_relation}), this gives
    \be
    S_{EE}(c) - \frac{\pi l(c+\bar c)}{48\epsilon}   = \ln S^C_{i0} = \ln d_i - \ln D_C. 
    \ee
    In other words, the {\it topological entanglement is dictated completely by the auxiliary phase $C$ characterizing the interface.} 
    
    What this tells us is that the basis state that we have chosen  ``resolves'' different topological sectors up to the resolution described by the phase $C$. As a result, the topological entanglement entropy is completely controlled by topological data of the auxiliary phase $C$. 
    
    In general where we have an interface obtained by fusion of many intermediate interfaces, we would have the choice of building up basis states that specify a particular topological sector in any one of the intermediate state, while other anyon lines matched to this intermediate segment are summed over as above. Then the entanglement entropy would be characterized purely by properties of this intermediate anyon line.

    We will work out some explicit examples in the following.
    
    \subsection{Example: entanglement across the $D(\bZ_M)-D(\bZ_N)$ interface}
    The Ishibashi state describing one $D(\bZ_M)-D(\bZ_N)$ interface has been constructed in equation (\ref{MN_ishibashi}). The boundary condition impose restrictions on quantum numbers $(r,s,(\beta),x,b)$ which determines the sector the anyon line pierces through the edge. Consider the case where two physical boundaries $B_1$ and $B_2$ are both characterized by the electric condensate, which implies $\beta=0$ and $x=0$. Consider for example the following state 
    \be
    |B\rangle\rangle=  | r(\beta=0),s(\beta=0),x=0,b\rangle\rangle
    \ee
    The reduced density matrix is obtained by tracing out the upper edge in the entanglement cut. Using (\ref{define_rsxb}) and the fact that $\chi_{r,s,x,b}$ are characters of $D(\bZ_{Cmn})$,  we immediately recover the entanglement entropy
    \be
    S_{EE}(r,s,x,b)=\frac{{\pi l}}{{12\epsilon}}-\ln(Cnm).
    \ee
    Supposedly one can take arbitrary linear combination of these basis states, and obtain extra contributions to the entanglement entropy analogous to the last term in (\ref{SEE_general}). These numbers would have appeared meaningless at first sight. 
    On the other hand, as we have already seen, there are special linear combinations where these numbers take upon a topological meaning reflecting enhanced chiral symmetry, or anyon condensation.  For example, for a state where we sum over $r,s$ independently, which is equivalent to summing $\beta$, we have
    \be
    |B'\rangle\rangle=\sum_{\substack{0\leq r \leq m-1, \\ 0\leq s\leq n-1}} |r,s,x=0,b\rangle\rangle,
    \ee
    which is a state that preserves the chiral symmetry algebra characterizing $D(\bZ_C)$.
    The entanglement of this state is given by
    \be
    S_{EE}(r,s,x,b)=\frac{{\pi l}}{{12\epsilon}}-\ln(C). 
    \ee

    \subsection{Example: entanglement across the $D(S_3)-D(\bZ_4)$ interface}
    
    In section \ref{43_interface} we described the construction of an interface between $D(S_3)-D(\bZ_4)$.
    Consider an interface controlled by an auxiliary phase $D(\bZ_2)$. i.e. Both $D(S_3)$ and $D(\bZ_4)$ condense to $D(\bZ_2)$. Then a natural basis state to consider (e.g. corresponding to the trivial sector 
    of $D(\bZ_2)$) would be given by
    \begin{align}
    |0e,0m\rangle_{D(\bZ_2)} = & |(A,(0e,0m))\rangle_{D(S_3) \otimes D(\bZ_4)} + |(A,(2e,0m))\rangle_{D(S_3) \otimes D(\bZ_4)}+ \\ \nonumber 
    & |(C,(0e,0m))\rangle_{D(S_3) \otimes D(\bZ_4)}+ |(C,(2e,0m))\rangle_{D(S_3) \otimes D(\bZ_4)}.
    \end{align}
    The entanglement across the interface for this state would be given by the quantum dimension of the auxiliary phase -- i.e. of $D(\bZ_2)$:
    \be
    S_{EE} - \frac{\pi l (c+\bar{c})}{48\epsilon} = -\ln 2.  
    \ee
    In the language of the edge CFT, we have chosen a combination of states that admits chiral symmetry enhancement at the boundary. 

    \section{Conclusion}
    
 In this paper, we explored the effect of gapped boundaries or interfaces on entanglement entropy. We first elaborated on the connection between anyon condensation and gapped boundaries/interfaces. Based on the physical data of anyon condensation, that gives a generic procedure to build a set of ground state basis. For each ground state basis, we demonstrate in examples how the conformal boundary conditions at the entanglement cut, appropriate for the gapless edge modes in the extended Hilbert space, is determined directly or influenced by anyon condensation at the interface/boundary. 
 
Particularly we considered two classes of cases. First, we consider cases where the entanglement cut is not in contact with the interfaces/boundaries, and second, where the cut coincides with the interface. In the first case, the interface determines the allowed anyon flux that can pass through the entanglement cut, whose quantum dimensions contribute to the entanglement entropy subsequently.  We show that previous work based on lattice gauge theories \cite{chen_entanglement_2018} can be reproduced using the current method.  
In the second case, the entanglement entropy depends on the extent of chiral symmetry breaking/enhancement at the interface -- this can be understood by the pattern of anyon condensation that defines the interface. The entanglement entropy is then determined by the quantum dimension of the ``auxiliary'' intermediate phase that describes the preserved chiral symmetry at the interface. We notice that chiral symmetry breaking at an interface is the predominant scenario considered in the CFT literature \cite{Fuchs:1992nq, petkova_generalised_2001}. In the context of topological phases however, it is perhaps more natural to consider chiral symmetry enhancement, which corresponds to breaking of the topological symmetry of the physical phases to a ``smaller'' auxiliary intermediate phase via anyon condensation.

These demonstrate the intricate interplay between anyon condensation and the structure of entanglement of the resultant ground states. 

In an accompanying paper \cite{We3}, we will consider the case where the entanglement cut passes through the interface or ends at the boundary. The methods and perspective discussed in this paper can be applied also in those cases, with the extra complication that careful treatment of the end point of the entanglement cut is needed.  More important data regarding anyon condensation pertaining in particular to the confined sector will be revealed in the entanglement. This contrasts with the analysis in the current paper where the condensed sector plays a major role.  We will report these interesting results there.

    \appendix
    
    \section{Setting the notations of Abelian Chern-Simons theories} \label{CSreview}
    
    The class of Abelian Chern-Simons theories that we are going to consider in the following takes the following form: 
    \begin{equation}
    S_{CS}=\frac{1}{4\pi}\int_M K^{IJ}A_I\wedge F_J, \qquad F_J = dA_J.
    \end{equation}
    Where $M$ is a 3d manifold. Here $K_{IJ}$ is a symmetric integral matrix and $I=1,...,N$. 
    Quantization would involve gauge fixing (such as taking the temporal gauge $A^I_t=0$) and solving for the constraints following from the gauge choice. A review of its detailed procedure can be found for example in \cite{wang_boundary_2015}. Upon gauge fixing, the action becomes a total derivative. In the temporal gauge for example, the constraint equation would amount to the flat condition 
    \be F_{I\,\,xy} = 0. \ee 
    Setting 
    \be A_{I\,\,x,y} =\partial_{x,y} \Phi_I \ee 
    for some scalar function $\Phi$ and substituting these expressions into the bulk action, we recover a total derivative term. 
    For $M$ an open manifold with a 2d boundary $\partial M$, the total derivative gives rise to the following boundary action 
    \begin{equation} \label{CSaction}
    S_{\partial M}=\frac{{1}}{{4\pi}}\int_{\partial M} dtdx\,(K^{IJ}\partial_t\Phi_I\partial_x\Phi_J-V^{IJ}\partial_x\Phi_I\partial_x\Phi_J),
    \end{equation}
    There is an extra term involving an integral symmetric matrix $V^{IJ}$  of rank $m$. As discussed in \cite{Wen:1995qn}, it is not determined by the bulk CS action. They can be viewed as physical parameters that depend on the actual material supporting these gapless edge modes.  
    We note that $m$ being even is a necessary (although not sufficient) condition for the edge modes to be ``gappable'' by relevant perturbation. 
    The boundary action can be quantized canonically.
    This gives, at constant time $t$,
    \be
    [\Phi_I(x),\Pi^J(y)] = i \delta^J_I\delta(x-y), \qquad \Pi^I(x)=\frac{{1}}{{2\pi}}K^{IJ}\partial_x\Phi_J.
    \ee 
    Assuming that $x$ is compact and that $x \sim x+ l$ i.e. the boundary at constant time $t$ is a ring of length $l$. The mode expansion of $\Phi_I$ at $t=0$ is given by
    \begin{equation}
    \label{eq:ModeExpansion}
    \Phi_I(x)=\phi_{0I}+K_{IJ}^{-1}P^{J}\frac{{2\pi}}{{l}}x+i\sum_{n\neq 0}\frac{{1}}{{n}}a_{I,n}e^{-inx\frac{{2\pi}}{{l}}}
    \end{equation}
    These modes therefore satisfy
    \be
    [\alpha_{I,n},\alpha_{J,m}]=nK_{IJ}^{-1}\delta_{n,-m}, 
    \ee 
    and for zero modes we have:
    $[\phi_{0I},P^{J}]=i\delta^{J}_I$.
    
    We will focus on the Chern-Simons equivalence of the $D(Z_N)$ models in the following.  The corresponding $K$ matrix is given by
    \be \label{kmatrixZN}
    K= \begin{pmatrix}
    0&N\\N&0
    \end{pmatrix}.
    \ee
    The matrix has a pair of eigenvalues with opposite sign, signifying that it has exactly one pair of left and right moving modes, and as such,  is a non-chiral theory. 
    The scalars $\Phi_I$ are related to the left and right moving fields by
    \begin{equation} \label{left_rightmodes}
    \Phi_1=\sqrt{\frac{{r}}{{2N}}}(\Phi_L+\Phi_R), \qquad
    \Phi_2=\sqrt{\frac{{1}}{{2Nr}}}(\Phi_L-\Phi_R)
    \end{equation}
    Here $r^2=V^{22}/V^{11}$.
    The left and right moving modes can also be expressed in a mode expansion:
    \be \label{expand_LR}
     \Phi_{L(R)}(x)=\phi_{0L(R)}+P_{L(R)}\frac{{2\pi}}{{l}}x+i\sum_{n\neq 0}\frac{{1}}{{n}}\alpha_{L(R),n}e^{-inx\frac{{2\pi}}{{l}}}, \ee
     To avoid clutter, we will take $r=1$ in the following. We note that $r$ does not play any role in the topological entanglement of a single non-chiral phase. One can show that it is canceled out in the computation of the topological entanglement. It does play a non-trivial role in the discussion of generic interfaces between different $D(Z_N)$ theories. We will re-introduce them where necessary.  We also note that when discussing topological entanglement in a chiral phase, one needs particular care in the choice of $r$. A detailed discussion will be taken up in the accompany paper. In that case, the entanglement cut crosses the physical interfaces, and extra care is needed. 
    Using the commutation relations of $\Phi_I$, we recover
    \be \label{quantized_P}
    [\alpha_{L(R),n},\alpha_{L(R),m}]=n\delta_{n,-m}, \ee and
    \be \label{left_right_momenta}
     P_{L,R}=\frac{{1}}{{\sqrt{2N}}}(\pm P^{1}+P^{2}). \ee
    The $U(1)$'s gauge groups are taken to be compact. Therefore the scalars are also compact, satisfying
    \be 
    \Phi_I \sim \Phi_I + 2\pi.
    \ee
    The conjugate momenta to the zero modes therefore are quantized, satisfying
    \be \label{quantized}
    P^{I}\in\mathbb{Z}.
    \ee
    We note that these $P_{\Phi_I}$ parametrizes a set of highest weight states. One can identify these highest weight states/operators with distinct anyons of the quantum double $D(Z_N)$.  The identification with anyons is many-to-one: $P^{I}$ and $P^{I}+ N $ describe the same topological sector. One can take $P^{1} \mod N$ to parametrize the electric charge wrt to the $Z_N$ gauge group in $D(Z_N)$ models, and $P^{2}$ the magnetic charges. A detailed review can be found in \cite{wang_boundary_2015}. We only record the basic set of facts needed in the current paper.   
    The Hamiltonian is given by
    \begin{equation} \label{Hamiltonian}
    H=\frac{{1}}{{4\pi}}\int_0^l dx(\partial_x\Phi_L \partial_x\Phi_L+\partial_x\Phi_R\partial_x\Phi_R)=\frac{{P_L^2+P_R^2}}{{2}}+\sum_{n>0}(\alpha_{L,-n}\alpha_{L,n}+\alpha_{R,-n}\alpha_{R,n})-\frac{{1}}{{12}}
    \end{equation}

    \section{Some useful details of the $D(S_3)$ model} \label{S3model_review}
    We would like to review here some basic data of the $D(S_3)$ model. 
    The anyons are labeled by $(A,r_{\alpha_A})$, where $A$ is a conjugacy class of the group $G=S_3 = \langle x,y\vert x^3=y^2= xyxy= e\rangle $ corresponding to magnetic charges, and $r_{\alpha_A}$ an irrep of the centralizer of $A$, corresponding to electric charges. A summary of all the anyons are listed below. 
    
        \begin{tabular}{c|ccc|cc|ccc}
            \hline\hline
             & $A$ & $B$ & $C$ & $D$ & $E$ & $F$ & $G$ & $H$ \\
            \hline
            conjugacy class $W$ & \multicolumn{3}{|c|}{$\{e\}$} & \multicolumn{2}{c|}{$\{y,xy,x^2y\}$} & \multicolumn{3}{c}{$\{x,x^2\}$} \\
            \hline
            centralizer $\cong$ & \multicolumn{3}{|c|}{$S_3$} & \multicolumn{2}{c|}{$Z_2$} & \multicolumn{3}{c}{$Z_3$} \\
            \hline
            irrep $\rho$ of centralizer  & \boldmath$1$ & sign & \boldmath$\pi$ & \boldmath$1$ & \boldmath$-1$ & \boldmath$1$ & \boldmath$\omega$ & \boldmath$\omega^*$ \\
            \hline
            dim($\rho$) & 1 & 1 & 2 & 1 & 1 & 1 & 1 & 1 \\
            \hline
            quantum dimension $d=|W|\times$ dim$(\rho)$ & 1 & 1 & 2 & 3 & 3 & 2 & 2 & 2\\
            \hline
            twist $\theta$ & 1 & 1 & 1 & 1 & -1 & 1 & $e^{2\pi i/3}$ & $e^{-2\pi i/3}$\\
            \hline\hline
        \end{tabular} 
    
    Their fusion rules are given by
    
        \scalebox{0.7}{
        \begin{tabular}{c|ccc|cc|ccc}
            \hline\hline
            $\otimes$ & $A$ & $B$ & $C$ & $D$ & $E$ & $F$ & $G$ & $H$ \\ \hline
            $A$ & $A$ & $B$ & $C$ & $D$ & $E$ & $F$ & $G$ & $H$ \\ 
            $B$ & $B$ & $A$ & $C$ & $E$ & $D$ & $F$ & $G$ & $H$ \\ 
            $C$ & $C$ & $C$ & $A\oplus B\oplus C$ & $D\oplus E$ & $D\oplus E$ & $G\oplus H$ & $F\oplus H$ & $F\oplus G$ \\ \hline
            $D$ & $D$ & $E$ & $D\oplus E$ & $A\oplus C\oplus F\oplus G\oplus H$ & $B\oplus C\oplus F\oplus G\oplus H$ & $D\oplus E$ & $D\oplus E$ & $D\oplus E$ \\ 
            $E$ & $E$ & $D$ & $D\oplus E$ & $B\oplus C\oplus F\oplus G\oplus H$ & $A\oplus C\oplus F\oplus G\oplus H$ & $D\oplus E$ & $D\oplus E$ & $D\oplus E$ \\ \hline
            $F$ & $F$ & $F$ & $G\oplus H$ & $D\oplus E$ & $D\oplus E$ & $A\oplus B\oplus F$ & $C\oplus H$ & $C\oplus G$ \\ 
            $G$ & $G$ & $G$ & $F\oplus H$ & $D\oplus E$ & $D\oplus E$ & $C\oplus H$ & $A\oplus B\oplus G$ & $C\oplus F$ \\
            $H$ & $H$ & $H$ & $F\oplus G$ & $D\oplus E$ & $D\oplus E$ & $C\oplus G$ & $C\oplus F$ & $A\oplus B\oplus H$ \\
            \hline\hline
        \end{tabular}
        }

    The $S$-matrix is given by
    \begin{eqnarray}
        \label{eq:S3_S_matrix}
        S=\frac{1}{6}\left(
        \begin{array}{cccccccc}
           1 & 1 & 2 & 3 & 3 & 2 & 2 & 2 \\
           1 & 1 & 2 & -3 & -3 & 2 & 2 & 2 \\
           2 & 2 & 4 & 0 & 0 & -2 & -2 & -2 \\
           3 & -3 & 0 & 3 & -3 & 0 & 0 & 0 \\
           3 & -3 & 0 & -3 & 3 & 0 & 0 & 0 \\
           2 & 2 & -2 & 0 & 0 & 4 & -2 & -2 \\
           2 & 2 & -2 & 0 & 0 & -2 & -2 & 4 \\
           2 & 2 & -2 & 0 & 0 & -2 & 4 & -2 \\
        \end{array}    
        \right),
    \end{eqnarray} 
    
    In order to make comparison between the results in \cite{chen_entanglement_2018} and those in the current paper, we review here some basic details of the ribbon operators in a Kitaev lattice gauge realization of the quantum double model. 
    
    The lattice model is defined such that every link has a $|G|$ dimensional Hilbert space with basis vectors labeled by group elements $g\in G$. 
        For a ribbon $\xi$, let $F_{\xi}^{g,h}$ be the ribbon operator whose action is reviewed for example in \cite{beigi_quantum_2011}. By applying $F_{\xi}^{g,h}$ on the ground state, we generate a quasiparticle anti-quasiparticle pair at the end points of ribbon $\xi$. i.e. these end points cease to commute with the Hamiltonian.\cite{kitaev_fault-tolerant_2003} For a given ribbon $\xi$, there're $|G|\times |G|$ linearly independent ribbon operators forming a $|G|\times |G|$ dimensional ``ribbon space'' $\mathcal{F}_{\xi}=\{ \sum_{h,g\in G}^{}c_{h,g}F_{\xi}^{h,g} \}$. 
    
    %
        
    As it is noted in \cite{beigi_quantum_2011}, the gapped boundaries of the lattice model can be classified by subgroups $K\subset G$. Degrees of freedom lying at the boundary links for a given $K$-boundary are restricted to a Hilbert subspace, where the basis labels are now restricted to $k\in K$. Connection between these subgroups $K$ and anyon condensation for $G=S_3$ is described in detail in the main text.

     The connection to anyon condensation suggests that only  ribbon operators in a subspace of the ribbon space $\mathcal{F}_{\xi}$ can end at the boundary without energy cost.  
     Denote this subspace by $\mathcal{C}_{\xi}$, it is known that $\mathcal{C}_{\xi}$ satisfies the following two conditions\cite{beigi_quantum_2011}:
        \begin{eqnarray}
            c_{khk^{-1},kg}&=&c_{h,g}\ \forall\ k\in K, \nonumber\\
            c_{h,g}&=&0  \quad\  \forall \ h\notin K 
        \end{eqnarray}

        In our cylindrical case we have $K_1=\{e\}$ at top boundary, and $K_2=\{e,y\}$ at bottom boundary. For a ribbon connecting the two physical boundaries, the subspace $\mathcal{C}_{\xi}$ is subject to two sets of constraints. $K_1$ boundary condition indicates that only coefficients $\{c_{e,g}|g\in S_3\}$ are non-zero, and $K_2$ boundary condition gives matching rule $c_{e,g}=c_{e,gy} \quad \forall g\in S_3$. So we've found the basis of $\mathcal{C}_{\xi}$ as a 3 dimensional vector space
        \begin{eqnarray}
            \label{eq:basis}
            \mathcal{C}_{\xi}=\{\alpha(F^{e,e}+F^{e,y})+\beta(F^{e,x}+F^{e,xy})+\gamma(F^{e,x^2}+F^{e,x^2y}) \}
        \end{eqnarray} 
        
        Note that only anyon $A$ and $C$ may condense at both boundaries (Table \ref{tab:condense}), this means the Wilson line connecting the two boundaries can only fluctuate between the two topological sectors. Since $\rho_A$ and $\rho_C$ are respectively 1d and 2d irrep of $D(S_3)$ (also of $S_3$ because they're chargeons), we can identify $A$ and $C$ with their respective 1d and 2d invariant subspace in $C_{\xi}$ as following
        \begin{eqnarray}
            \label{eq:repBasis}
            \begin{cases}
                \ket{A}=\frac{1}{\sqrt{3}}\ket{F^{e,e}+F^{e,y}}+\frac{1}{\sqrt{3}}\ket{F^{e,x}+F^{e,xy}}+\frac{1}{\sqrt{3}}\ket{F^{e,x^2}+F^{e,x^2y}} \\
                \ket{C_1}=\frac{1}{\sqrt{6}}\ket{F^{e,e}+F^{e,y}}+\frac{1}{\sqrt{6}}\ket{F^{e,x}+F^{e,xy}}-\frac{2}{\sqrt{6}}\ket{F^{e,x^2}+F^{e,x^2y}} \\
                \ket{C_2}=\frac{1}{\sqrt{2}}\ket{F^{e,e}+F^{e,y}}-\frac{1}{\sqrt{2}}\ket{F^{e,x}+F^{e,xy}}
            \end{cases}
        \end{eqnarray}
        where we use $\ket{F^{e,e}+F^{e,y}}$ to denote the normalized state $(F^{e,e}+F^{e,y})\ket{\psi}$. The 1d space spans $\{\ket{A}\}$ is the representation space of $\rho_A$, and the 2d space spans $\{\ket{C_1},\ket{C_2}\}$ is the representation space of $\rho_C$. Note that the states $\ket{A}$, $\ket{C_1}$ and $\ket{C_2}$ are eigenstates of some magnetic ribbon operators winding around the non-contractible cycle.
     
        In \cite{chen_entanglement_2018} we choose $\ket{\psi}$ as the ground state. $\ket{\psi}=F^{e,e}\ket{\psi}$ since operator $F^{e,e}$ does nothing to the state. It is also invariant under the action of $F^{e,y}$ because of the element $y$ in the bottom boundary subgroup $K_2$, so it can be identified with $\ket{F^{e,e}+F^{e,y}}$ here:
        \[ \ket{\psi}=F^{e,e}\ket{\psi}=F^{e,y}\ket{\psi}=\ket{F^{e,e}+F^{e,y}}. \]
        Formula (\ref{eq:repBasis}) can be reversed to express the ground state using Wilson line basis:
        \begin{eqnarray}
                \ket{F^{e,e}+F^{e,y}}&=&\frac{1}{\sqrt{3}}\ket{A}+\frac{1}{\sqrt{6}}\ket{C_1}+\frac{1}{\sqrt{2}}\ket{C_2}\nonumber\\
                &=&\frac{1}{\sqrt{3}}\ket{A}+\sqrt{\frac{2}{3}}\ket{\tilde{C}},
        \end{eqnarray}
      we see that the Wilson line has probability $\frac{1}{3}$ in topological sector $A$ and $\frac{2}{3}$ in topological sector $C$. A direct calculation of the topological entanglement entropy using formula 
        \begin{eqnarray}
            \label{eq:annulus}
            S_{TEE} =-2\ln D+2\sum_{a}^{}|\psi_a|^2\ln d_a -\sum_a^{}|\psi_a|^2\ln |\psi_a|^2 
        \end{eqnarray}
        provided in \cite{wen_edge_2016} gives $S_{TEE}=-\frac{1}{3}\ln 2-\ln 6$, which is in complete agreement with our previous result \cite{chen_entanglement_2018}.

    
    \section{$\eta$ and $\theta$ functions}
    We list here the definitions and basic properties of Dedekind $\eta-$function and Jacobi $\theta-$function.
        \begin{eqnarray}
            \eta(\tau)=q^{\frac{1}{24}}\prod_{n=1}^{\infty}(1-q^n),
        \end{eqnarray}
        \begin{eqnarray}
            \theta_2(\tau)=\sum_{n\in\mathbb{Z}}^{}q^{\frac{1}{2}(n+\frac{1}{2})^2}&=&2\eta(\tau)q^{\frac{1}{12}}\prod_{r=1}^{\infty}(1+q^r)^2, \nonumber\\
            \theta_3(\tau)=\sum_{n\in\mathbb{Z}}^{}q^{\frac{n^2}{2}}&=&\eta(\tau)q^{-\frac{1}{24}}\prod_{r=0}^{\infty}(1+q^{r+\frac{1}{2}})^2, \nonumber\\
            \theta_4(\tau)=\sum_{n\in\mathbb{Z}}^{}(-1)^nq^{\frac{n^2}{2}}&=&\eta(\tau)q^{-\frac{1}{24}}\prod_{r=0}^{\infty}(1-q^{r+\frac{1}{2}})^2
        \end{eqnarray}
        where $\tau$ is the modular parameter and $q=e^{2\pi i \tau}$. 

        These functions are related by modular $T$ transformation ($\tau\to\tau+1$) and $S$ transformation ($\tau\to-\frac{1}{\tau}$):
        \begin{eqnarray}
            \eta(\tau+1)=e^{\frac{\pi i}{12}}\eta(\tau),& \qquad \eta(-\frac{1}{\tau})=\sqrt{-i\tau}\eta(\tau)\nonumber\\
            \theta_2(\tau+1)=e^{\frac{\pi i}{4}}\theta_2(\tau),&\qquad \theta_2(-\frac{1}{\tau})=\sqrt{-i\tau}\theta_4(\tau), \nonumber\\
            \theta_3(\tau+1)=\theta_4(\tau),&\qquad \theta_3(-\frac{1}{\tau})=\sqrt{-i\tau}\theta_3(\tau), \nonumber\\
            \theta_4(\tau+1)=\theta_3(\tau),&\qquad \theta_4(-\frac{1}{\tau})=\sqrt{-i\tau}\theta_2(\tau)
        \end{eqnarray}
        In the body $\tilde{q}$ is defined as the $S$ transformation of corresponding $q$, for any complex number $X$
        \begin{equation}
            q=e^X\xrightarrow{S}\tilde{q}=e^{\frac{4\pi^2}{X}}
        \end{equation}

    \section*{Acknowledgements}
    We thank Laurent Freidel, Gabriel Wong and Yang Zhou for helpful discussions. We would also like to thank Juven Wang for many past conversations on the subject.
    We are particularly grateful to Yidun Wan for a critical reading of our draft. 
    Part of this work is done during LYH and CS's visit to Perimeter Institute, as part of the Emmy-Noether Fellowship programme. 
    LYH acknowledges the support of Fudan University and the Thousands Young Talents Program. This work is supported by the NSFC grant number 11875111.
     \bibliography{ref}
    \bibliographystyle{utphys}

    \end{document}